\documentclass[preprint,superscriptaddress,nofootinbib,12pt]{revtex4-1}
\usepackage[english]{babel}
\usepackage{graphicx}
\usepackage[letterpaper,top=2cm,bottom=2cm,left=2cm,right=2cm,marginparwidth=1.25cm]{geometry}
\usepackage{braket}
\usepackage{tabularx}
\usepackage{float}
\usepackage{amsmath}
\usepackage{physics}
\usepackage{graphicx}
\usepackage{xcolor}
\usepackage{comment}
\usepackage{multirow}
\usepackage{caption}
\usepackage{subcaption}
\usepackage[colorlinks=true, allcolors=blue]{hyperref}
\usepackage{lipsum}

\begin{document}
\title{Quantum Information Measures in Quartic and Symmetric Potentials using perturbative approach}

\author{Vikash Kumar Ojha}
\email{vko@svnit.ac.in}
\affiliation{Department of Physics, Sardar Vallabhbhai National
Institute of Technology, Surat, 395 007, India.}
\author{Ramkumar Radhakrishnan}
\email{rradhak2@ncsu.edu}
\affiliation{Department of Physics, North Carolina State University, Raleigh, NC 27695, USA.}
\author{Mariyah Ughradar}
\email{ds22ph003@phy.svnit.ac.in}
\affiliation{Department of Physics, Sardar Vallabhbhai National
Institute of Technology, Surat, 395 007, India.}
\begin{abstract}
We analyze the Shannon and Fisher information measures for systems subjected to quartic and symmetric potential wells. The wave functions are obtained by solving the time-independent Schr\"{o}dinger equation, using aspects of perturbation theory. We examine how the information for various quantum states evolves with changes in the width of the potential well. For both potentials, the Shannon entropy decreases in position space and increases in momentum space as the width increases, maintaining a constant sum of entropies, consistent with Heisenberg's uncertainty principle. The Fisher information measure shows different behaviors for the two potentials: it remains nearly constant for the quartic potential. For the symmetric well potential, the Fisher information decreases in position space and increases in momentum space as localization in position space increases, also consistent with the analogue of  Heisenberg's uncertainty principle. Additionally, the Bialynicki-Birula–Mycielski inequality is evaluated across various cases and is confirmed to hold in each instance. 

\end{abstract}

\maketitle
\section{Introduction}\label{sec:introduction}
There is always great interest in studying information-theoretic measures for quantum mechanical systems \cite{qiang2007analytical,maireche2020modified,ikot2021nikiforov,edet2021shannon,yu2004exactly,vitoria2018interaction,zhang2010exactly}. It is not only because of its foundational aspects but also due to its phenomenological relevance \cite{chao1972application,liu20192d}. We can use this as a first step to understand complex models or processes (see refs. \cite{hassanabadi2019analysis,das2010thermodynamics,bensalem2019statistical}). Among these are theoretical evaluations of information entropies for various potentials and systems, such as the Pöschl-Teller potential \cite{atre2004quantum}, the hyperbolic potential \cite{valencia2015quantum}, squared tangent potential \cite{dong2014quantum}, heavy mesons influenced by point-like defects \cite{almeida2023quantum}, symmetrically trigonometric Rosen Morse potential \cite{sun2013quantum}, as well as systems involving position dependent mass distributions  \cite{lima2022quantum,yanez2014quantum,guo2015shannon,falaye2016fisher,sun2015shannon,santana2022quantum}, and for nonlinear systems \cite{yamano2024shannon,ballesteros2023shannon,deffner2022nonlinear,song2015shannon}. Generally, the method chosen to solve the Schr\"{o}dinger equation depends on the specific quantum mechanical system, particularly when seeking an analytical solution. Various techniques exist to derive analytical expressions for such systems, including asymptotic iteration method (AIM) \cite{champion2008asymptotic,ismail2020asymptotic,ciftci2003asymptotic}, Laplace transformation method (LTM) \cite{schiff2013laplace,podlubny1997laplace,wilcox1978numerical}, Nikiforov-Uvarov method (NUM) \cite{yacsuk2005exact,berkdemir2012application,karayer2015extension} etc. However, finding analytical solutions is not always feasible for all quantum mechanical systems. In such cases, we rely on approximation techniques to solve the equation, providing practical solutions when exact methods are either too complex or impossible to apply. Some of the most powerful methods include the WKB approximation \cite{miller1953wkb,karnakov2012wkb} and perturbation theory \cite{capitani2003lattice,stevenson1981optimized,kato2013perturbation}. Perturbation theory, when applied with appropriate corrections, provides results that closely approximate the exact analytical solution, making it particularly useful for systems where small deviations from a known solution are considered. In our work, we obtained the wave function using perturbation theory with appropriate corrections.

Entropy, a measurable property of physical systems, is often associated with disorder, randomness, or uncertainty. In thermodynamics \cite{thermo}, entropy measures the degree of irreversibility in a physical system. However, it's important to distinguish this from information entropy, which is our primary focus here. Key differences exist between thermodynamic entropy and information entropy: thermodynamic entropy is concerned with the number of potential structural configurations in a system, while information entropy deals with the choices made in a communication system. There has been a growing interest in using concepts from information theory in the study of quantum systems. For example, entropic uncertainty relations are being investigated as potential alternatives to the Heisenberg uncertainty principle \cite{heisenberg}. Shannon entropy \cite{shannon} and these entropic uncertainty relations are being applied across various fields, including the analysis of squeezed states \cite{squeezing}, fractional revivals \cite{fractional, fractional1}, and the use of maximum entropy techniques to reconstruct charge and momentum distributions in atomic and molecular systems \cite{maximumentro}. Additionally, Shannon entropy \cite{shannon} finds applications in measure theory \cite{measure}, the study of open quantum systems with Markov chains \cite{markov}, decision tree methods in machine learning, and Bayesian inference \cite{bayesian}. Fisher's information measure, sometimes called Fisher's entropy, is an important concept originating from communication theory, and it serves as an early foundation for Shannon entropy \cite{fentro}. Fisher information captures how much information an observable variable contains about an unknown parameter of interest. In quantum mechanics, Fisher information plays a significant role in understanding measurement uncertainty \cite{uncefisher}. Measurements in quantum systems are inherently uncertain due to the probabilistic nature of quantum states, and Fisher's information helps quantify this uncertainty. It effectively measures how sensitive the probability of observed data is to changes in the parameter being measured. In other words, a higher Fisher information value implies that even small changes in the parameter lead to significant changes in the observed data, thus allowing for more precise estimations of the parameter. This concept provides insights into the accuracy and reliability of measurements in quantum systems and links to the broader field of information theory. Therefore, Fisher information is not just a tool for assessing uncertainty; it's a fundamental measure that bridges the gap between classical and quantum information theory, influencing how we understand and interpret data in complex systems \cite{fentro}. In our work, we have calculated the Shannon and Fisher information measures for two types of potentials, namely the quartic and symmetric potentials. These calculations were performed within the framework of perturbation theory, allowing us to explore the information-theoretic properties of these systems in an approximate analytical context. The double well potential also known as the quartic potential, holds significant importance in quantum mechanics, quantum field theory, and other other fields for investigating different physical phenomena and their properties. The symmetric well potential is ubiquitous as it serves as a model to illustrate the concept of instantons \cite{robo} and Feynman path integral formulation in quantum mechanics \cite{shankar}. Quantum systems in quartic potential are formed by the interplay between self-phase modulation and anomalous second- and fourth-order dispersion \cite{new}. This has a widespread application in the field of optical communication \cite{10,11}, such as shape in variance in nonlinear wave packets which enhances optical communication.

This article is arranged in the following way: In Section \ref{sec:description}, we briefly discuss the systems we have considered. In Section \ref{sec:information}, we derive the Shannon and Fisher information measures for these systems and then we conclude with Section \ref{sec:conclusion}.  All the calculations are carried out for systems in different states, using the convention $\hbar = c = K_{B} =  1$.
\section{Description of the system} \label{sec:description}
In this section, we briefly examine two distinct potentials and derive their respective wave functions. These potentials include the quartic potential, a form of double-well potential, and the symmetric potential. The examination of quartic potential is widespread because it demonstrates the concept of instantons in quantum field theories \cite{robo} and path integrals formulation in quantum mechanics \cite{shankar}. It also plays a significant role in analyzing the shape variations of wave packets in optical communications \cite{10,11,22}. Meanwhile, the symmetric well potential is crucial for managing dispersion in optical and quantum communication \cite{sympot}. So we begin by deriving the wave function associated with the quartic potential $V(x)$ defined as follows
\begin{equation}\label{eq:quarticpot}
    V(x) = ax^{2}+bx^{4}, b>0.
\end{equation}
\begin{figure}[H]
\begin{subfigure}{0.45\textwidth}
  \centering
  \includegraphics[width=\linewidth]{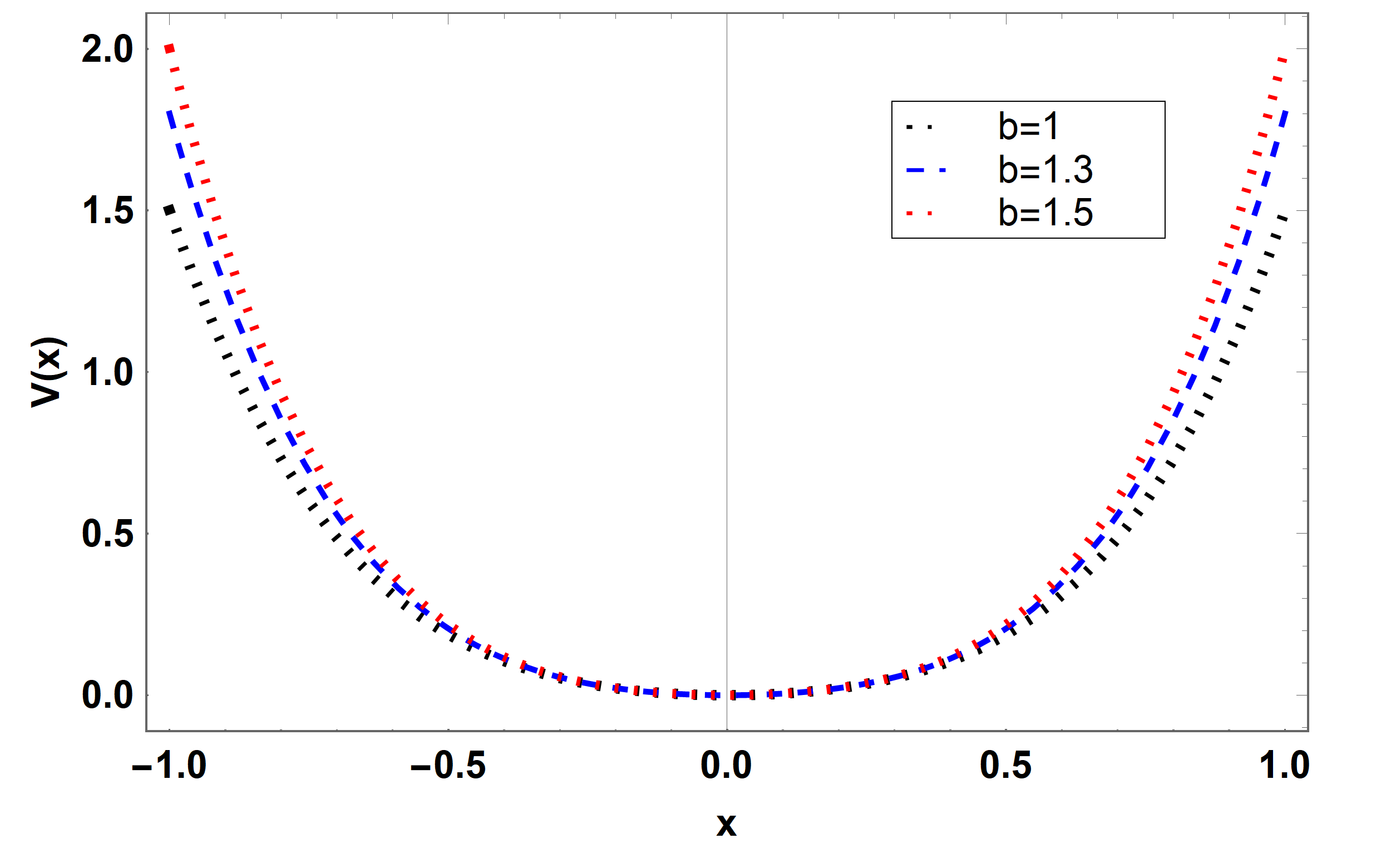}
  \label{fig:image123}
\end{subfigure}
\hspace{4em}
\begin{subfigure}{0.45\textwidth}
  \centering
  \includegraphics[width=\linewidth]{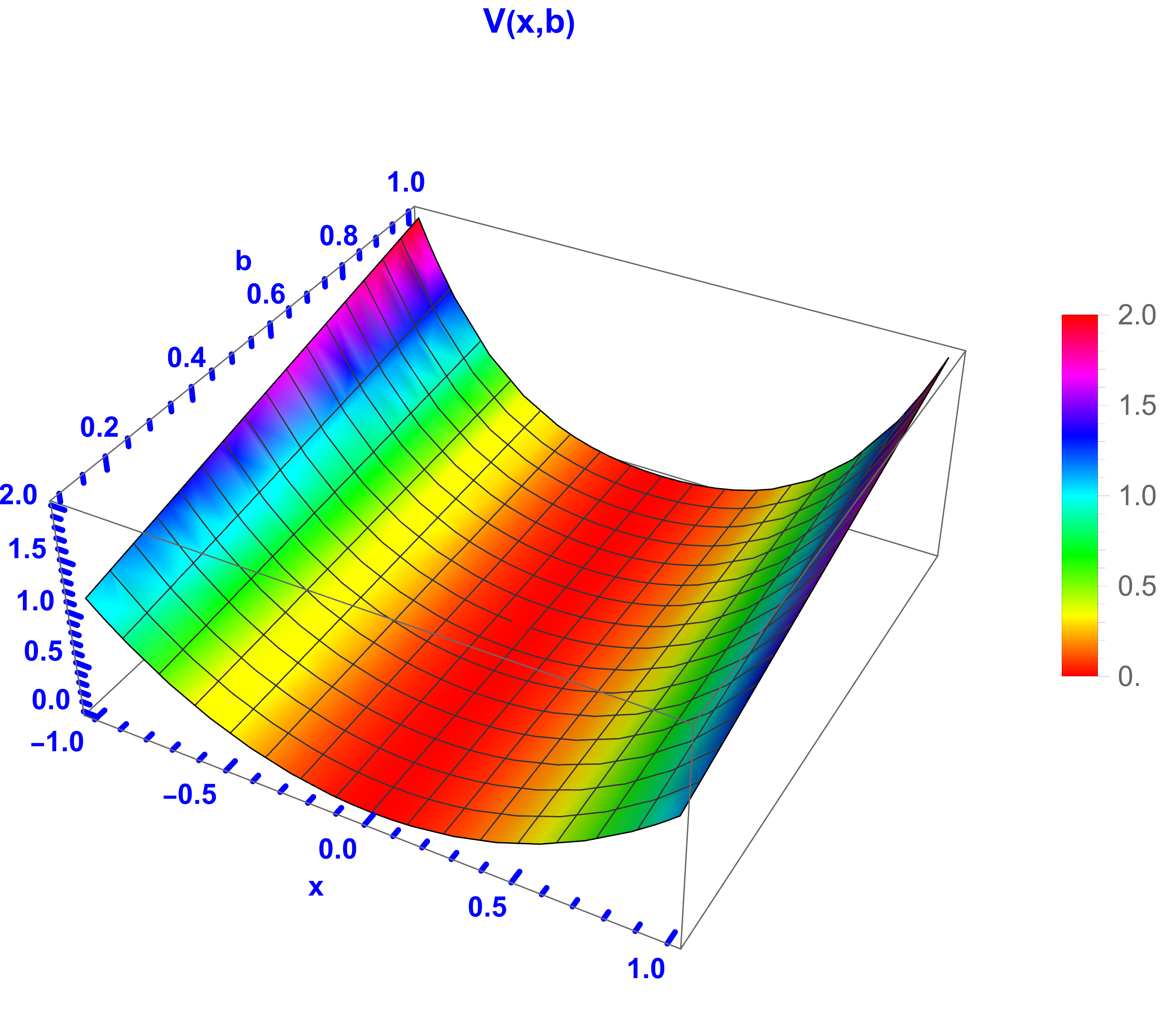}
   \label{fig:image2}
\end{subfigure}
 \caption{Plot of quartic potential for various values of the parameter $b$.}
 \label{fig:pot1}
\end{figure}
Schr\"{o}dinger equation for this system is expressed in a simplified form as:
\begin{equation}\label{eq:constmass}
 H\psi(x,t) = \frac{-1}{2}\frac{d^2\psi(x,t)}{dx^2}+\bigg(ax^2+bx^4\bigg) \psi (x,t).
\end{equation}
Throughout the article, we consider solutions that are time-independent, localized, and have finite energy. As a result, the time dependence of the wave functions is eliminated. Within the framework of quantum field theory, these characteristics can be described as solitonic solutions \cite{robo}. For simplicity, we set the value of the parameter $a$ to $1/2$. With these modification, equation \eqref{eq:constmass} becomes
\begin{equation}\label{eq:constantmass}
    H\psi(x) = \frac{-1}{2}\frac{d^2\psi(x)}{dx^2}+\bigg(\frac{x^2}{2}+bx^4\bigg) \psi (x),\hspace{0.5em} b>0.
\end{equation}
The analytical solution of the wave function is provided in \cite{dong2019exact,child2000quantum}. However, finding the Fourier component of it and calculating entropies would be cumbersome using that. So, we briefly apply the concept of time-independent perturbation theory to find the energy levels and wave functions of this system \cite{shankar}. To do this, we first consider the unperturbed potential $x^{2}$ and treat $bx^{4}$ as a perturbation to it. Using it, we then calculate the energy levels and wave functions\footnote{Throughout the article, subscripts are used to indicate energy levels, wave functions, and states, while superscripts represent the order of perturbation.}. The ground state wave function $(\psi^{(0)}_{n=0} (x))$ of this system is given by
\begin{equation}
    \psi^{(0)}_{n=0} (x) = \psi^{\prime (0)}_{n=0}(x) + \psi^{\prime (1)}_{n=0}(x)
\end{equation}
where $\psi^{\prime(0)}_{n=0}(x)$ represents the wave function without any perturbation, $\psi^{\prime(1)}_{n=0}(x)$ represents the wave function with the $bx^{4}$ perturbation and $ \psi_{n=0}^{\prime(1)}(x) = \sum_{m\neq n}\ket{m^{(0)}}\frac{\bra{m^{(0)}}bx^{4}\ket{n^{(0)}}}{E_{n}^{(0)}-E_{m}^{(0)}}$. The ground state wave function is given by
\begin{equation}
 \psi_{n=0} (x) =    -b\Bigg[\sqrt{\frac{3}{2}}\ket{4}+3\sqrt{2}\ket{2}\Bigg],
\end{equation}
where $\ket{4} = \frac{1}{\sqrt{16.4!}}\bigg(\frac{1}{\pi}\bigg)^{1/4}H_{4}(x)e^{\frac{-x^2}{2}}$, $\ket{2} = \frac{1}{\sqrt{4.2!}}\bigg(\frac{1}{\pi}\bigg)^{1/4}H_{2}(x)e^{\frac{-x^2}{2}} $ and $H_{n}$ denote Hermite polynomials \cite{andrews1999hermite}. Upon substitution and simplification, we obtain the normalized wave function in position space as
\begin{equation}
    \psi_{n=0} (x) = A_{0}\Bigg[1-b\bigg(\frac{1}{\sqrt{16.4!}}\sqrt{\frac{3}{2}}(16x^4-48x^2+12)+\frac{1}{\sqrt{4.2!}}3\sqrt{2}(4x^2-2)\bigg)\Bigg]e^{\frac{-x^2}{2}},
\end{equation}
where $A_{0} = \frac{1}{\sqrt{1+6.681\times 10^{-16}b+18.5b^{2}}}\bigg(\frac{1}{\pi}\bigg)^{1/4}$ is the normalisation constant. In Fourier space, the ground state wave function is given by
\begin{equation}
 \phi_{n=0} (k) = B_{0}\Bigg[12+b\bigg(-3(12+\sqrt{3})+12(6+\sqrt{3})k^{2}-4\sqrt{3}k^{4}\bigg)\Bigg]e^{\frac{-k^2}{2}},
\end{equation}
where $B_{0} = \frac{0.0692}{\sqrt{1+6.681\times 10^{-16}b+18.5b^{2}}}$ is the normalisation constant. We use the same principle to obtain the normalized wave function for the excited states. The wave functions in position $(x)$ and momentum $(k)$ space for the higher order states are as follows
\begin{equation}
    \psi_{n=1} (x) = A_{1}\Bigg[1-\frac{b}{32}\bigg\{10(2x^{2}-3)+\frac{1}{2}(4x^{4}-20x^{2}+15)\bigg\}\Bigg]xe^{-\frac{x^2}{2}},
\end{equation}
where $A_{1} = \frac{\sqrt{2}}{\pi^{1/4}\sqrt{1+2.505\times 10^{-16}b+0.6152 b^{2}}}$. 
In Fourier space
\begin{equation}
    \phi_{n=1}(k) = B_{1}\Bigg[4k^{4}b-60k^{2}b+75 b-64\Bigg]k e^{-\frac{k^{2}}{2}},
\end{equation}
where $B_{1} = \frac{-0.0165 i}{\sqrt{1+2.505\times 10^{-16}b+0.6152 b^{2}}}\sqrt{\frac{1.625 +4.071\times 10^{-16}b+b^{2}}{1.606+8.049\times 10^{-16}b + 0.9883b^{2}}}$. For $n=2$ state
\begin{equation}
 \psi_{n=2} (x) = A_{2}\frac{b}{16}\bigg(\frac{1}{\pi}\bigg)^{\frac{1}{4}}\bigg[12\sqrt{2}-14\sqrt{2}(4x^4-12x^2+3)-\frac{1}{\sqrt{2}}(8x^6-60x^4-90x^2-15)\bigg]e^{-\frac{x^2}{2}},
\end{equation}
where $A_{2} = \sqrt{\frac{128}{12183}} $. In momentum space
\begin{equation}
    \phi_{n=2}(k) = B_{2} e^{\frac{-k^2}{2}}(105 + 246 k^2 - 172 k^4 + 8 k^6)
\end{equation}
where $B_{2} = \frac{b}{2\sqrt{12183}\hspace{0.2em}\pi^{1/4}} $.For $n=3$ state
\begin{equation}
    \psi_{n=3} (x) = A_{3}\frac{b}{16}\bigg(\frac{1}{\pi}\bigg)^{\frac{1}{4}}\bigg[20\sqrt{3}-36\sqrt{5}(4x^5-20x^3+15)-\sqrt{\frac{2}{3}}(8x^7-84x^5+210x^3-105x)\bigg]e^{-\frac{x^2}{2}},
\end{equation}
where $A_{3} = \sqrt{\frac{-32}{45(-7163 + 60 \sqrt{15})}} $. In its corresponding Fourier space
\begin{equation}
    \phi_{n=3}(k) = B_{3} e^{\frac{-k^2}{2}}\bigg(60 (\sqrt{3} - 27 \sqrt{5}) - 105 i \sqrt{6} k + 
  30 i(72 \sqrt{5} + 7 \sqrt{6}) k^3 - 
  12 i (36 \sqrt{5} + 7 \sqrt{6}) k^5 + 8 i \sqrt{6} k^7\bigg)
\end{equation}
where $B_{3} = \frac{b}{18 \sqrt{10(7163-60\sqrt{15})}\pi^{\frac{1}{4}}} $. We have thus obtained the wave functions for both the ground state and higher excited states of the system under a quartic potential (Refer appendix \ref{sec:appendix1} for more details). In the next section, we will use these results to calculate entropies and other information measures. Now, we turn to deriving the wave functions for a system in a symmetric well. We will apply the principle of stationary perturbation theory, as we did for the quartic potential, to obtain the wave functions for this system. The potential corresponding to this is as follows
\begin{equation}
    V(x) = V_{0}\bigg(\frac{1-\lambda x \cot(\lambda x)}{(\lambda x)^{2}}\bigg),\hspace{1em} \lambda > 0,
\end{equation}

where $V_{0}$ is a constant and $\lambda$ is a parameter which will be varied. The time-independent Schr\"{o}dinger equation is given by
\begin{equation}
    H\psi = \frac{-1}{2}\frac{d^2\psi(x)}{dx^2} + V_{0}\bigg(\frac{1-\lambda x \cot(\lambda x)}{(\lambda x)^{2}}\bigg) \psi (x).
\end{equation}
    \begin{figure}[H]
        \centering
        \includegraphics[width=0.5\linewidth]{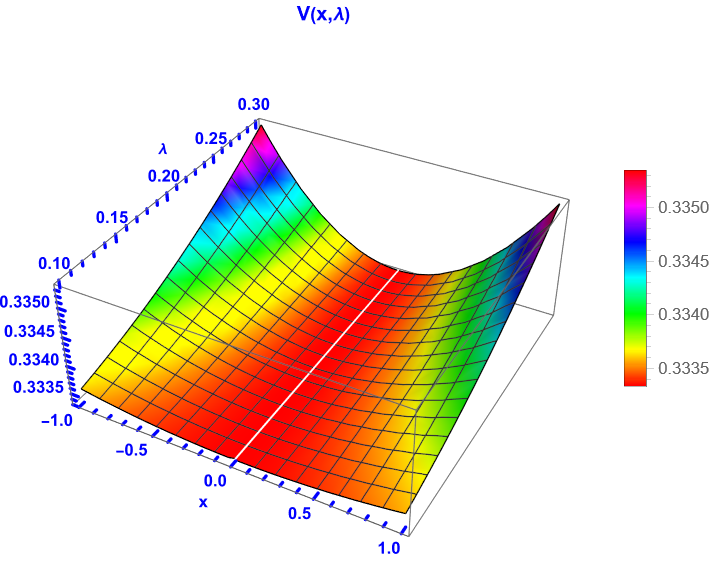}
        \caption{Plot of symmetric well for various parameter values $\lambda$.}
        \label{fig:potsym}
    \end{figure}

Unlike the quartic potential case, we do not provide the explicit values of the wave functions beyond $n=1$ here due to their considerable length. Instead, we present their calculations in appendix \ref{sec:appendix}. The ground state wave function $(n=0)$ in position space for a particle in a constant mass under a symmetric well is given by
\begin{equation}
    \psi_{n=0}(x) = \bigg(\frac{1}{\pi}\sqrt{\frac{2}{45}}\bigg)^{1/4} \Bigg[1-0.1694\bigg(\frac{2\sqrt{2}}{\sqrt{45}}\lambda^{2}x^{2}-1\bigg)-0.014\bigg(\frac{8}{45}\lambda^{4}x^{4}-12\sqrt{\frac{2}{45}}\lambda^{2}x^{2}+3\bigg)\Bigg]e^{\frac{-\lambda^{2}x^{2}}{3\sqrt{10}}}.
\end{equation}
In reciprocal space
\begin{equation}
     \phi_{n=0}(k) \sim \frac{0.4938}{\lambda^{9/2}}\bigg[1.717\lambda^{4}+5.235\lambda^{2}k^{2}-2.744k^{4}\bigg]e^{\frac{-3\sqrt{5}k^{2}}{2\sqrt{2}\lambda^{2}}}.
\end{equation}
Similarly the wave function in position space for the first excited state $(n=1)$ is given by
\begin{equation}
  \psi_{n=1}(x) \sim C_{1} \bigg[0.508967 - 0.177886 x \lambda - 0.0632295 x^3\lambda^3 -  0.00580632 x^5 \lambda^5\bigg]e^{\frac{-x^2\lambda^2}{3\sqrt{10}}}
\end{equation}
where $C_{1} = \frac{172.22}{\sqrt{\frac{1.275\times 10^{-12}}{\lambda}}+\frac{7.73 \times 10^{-12}}{\sqrt{\lambda^{2}}}}$ is the normalisation constant. In Fourier space
\begin{equation}
     \phi_{n=1}(k) \sim D_{1} \bigg[- 30.3670i k^5 + 78.7167i k^3 - 31.3778i k + 1.1084\bigg]e^{\frac{-3\sqrt{\frac{5}{2}}k^2}{2\lambda^2}},
\end{equation}
where $D_{1} = \frac{0.9997}{\sqrt{0.999 \lambda + 84.46 \lambda^{3}-134.007 \lambda^{5}+115.838 \lambda^{7}-50.435 \lambda^{9}+9.227\lambda^{11}}}$ is the normalisation constant. Similarly, all other wave functions obtained, including those for the $n=2,3$ states, are also normalized to unity. 
\begin{figure}[H]
\begin{subfigure}{0.5\textwidth}
  \centering
  \includegraphics[width=\linewidth]{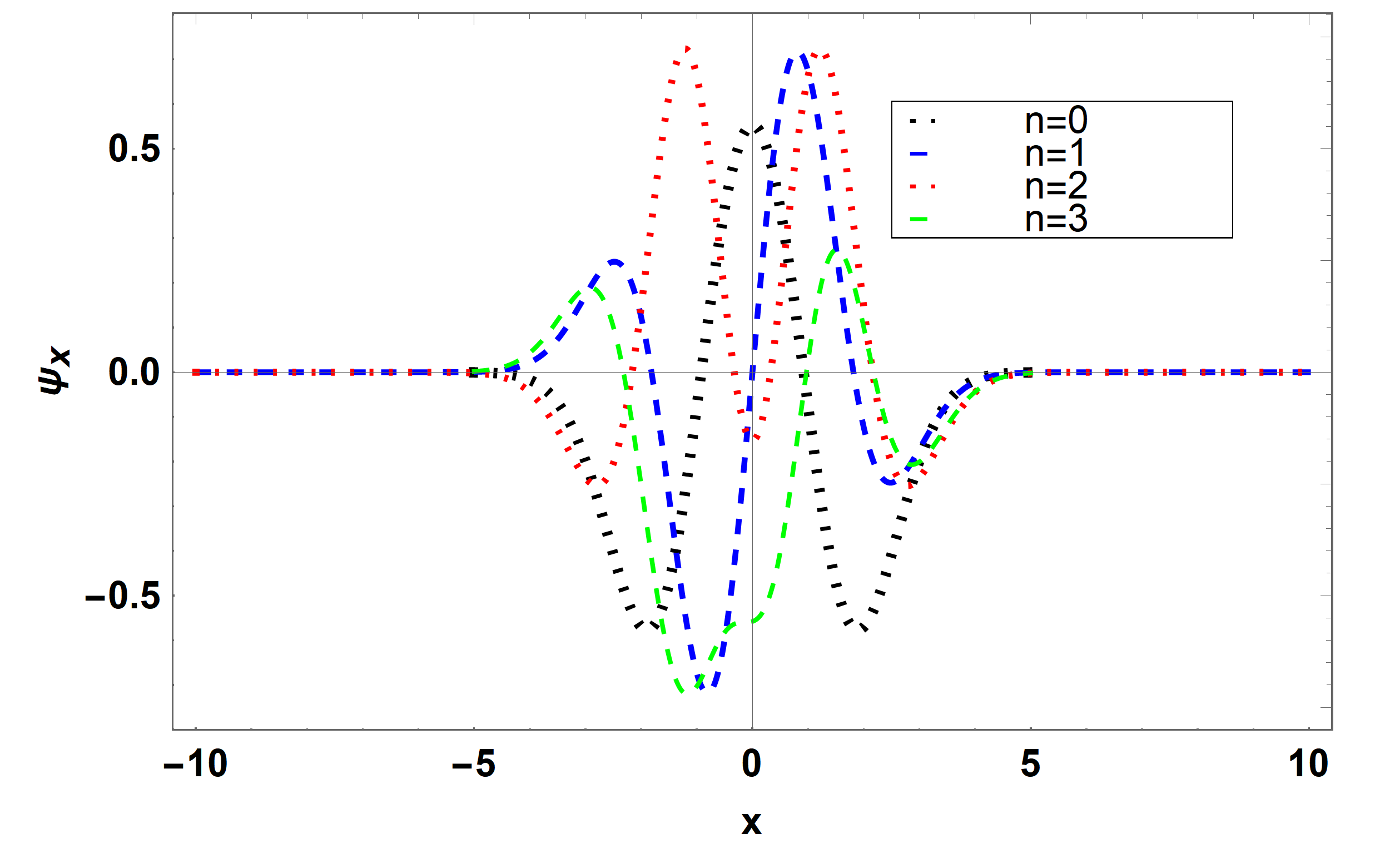}
  \caption{Position space}
  \label{fig:image111}
\end{subfigure}
\begin{subfigure}{0.5\textwidth}
  \centering
  \includegraphics[width=\linewidth]{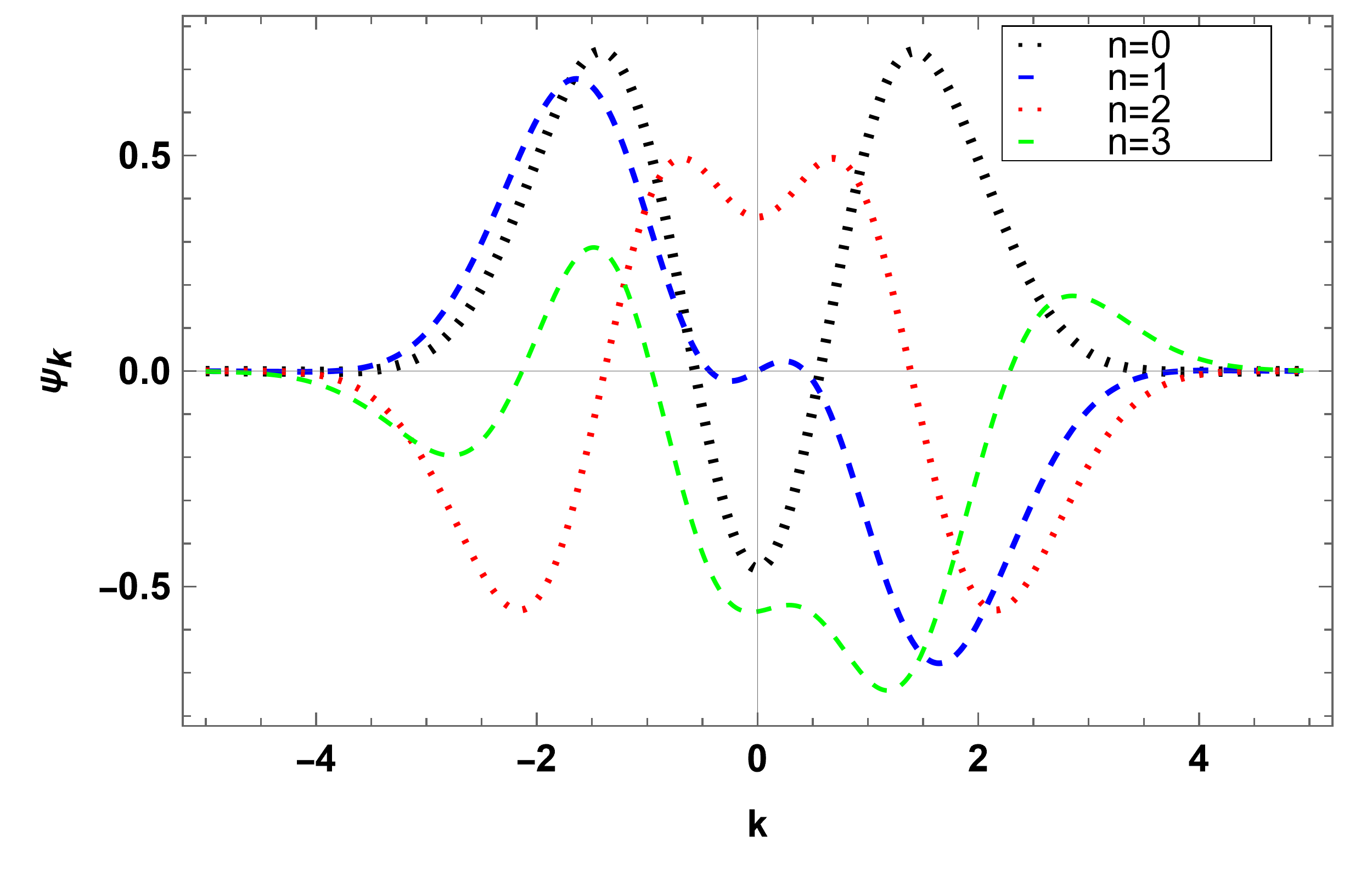}
  \caption{Momentum space}
   \label{fig:image26}
\end{subfigure}
 \caption{Wave functions of Quartic potential for various values of $n$. }
 \label{fig:wavef1}
\end{figure}

\begin{figure}[H]
\begin{subfigure}{0.5\textwidth}
  \centering
  \includegraphics[width=\linewidth]{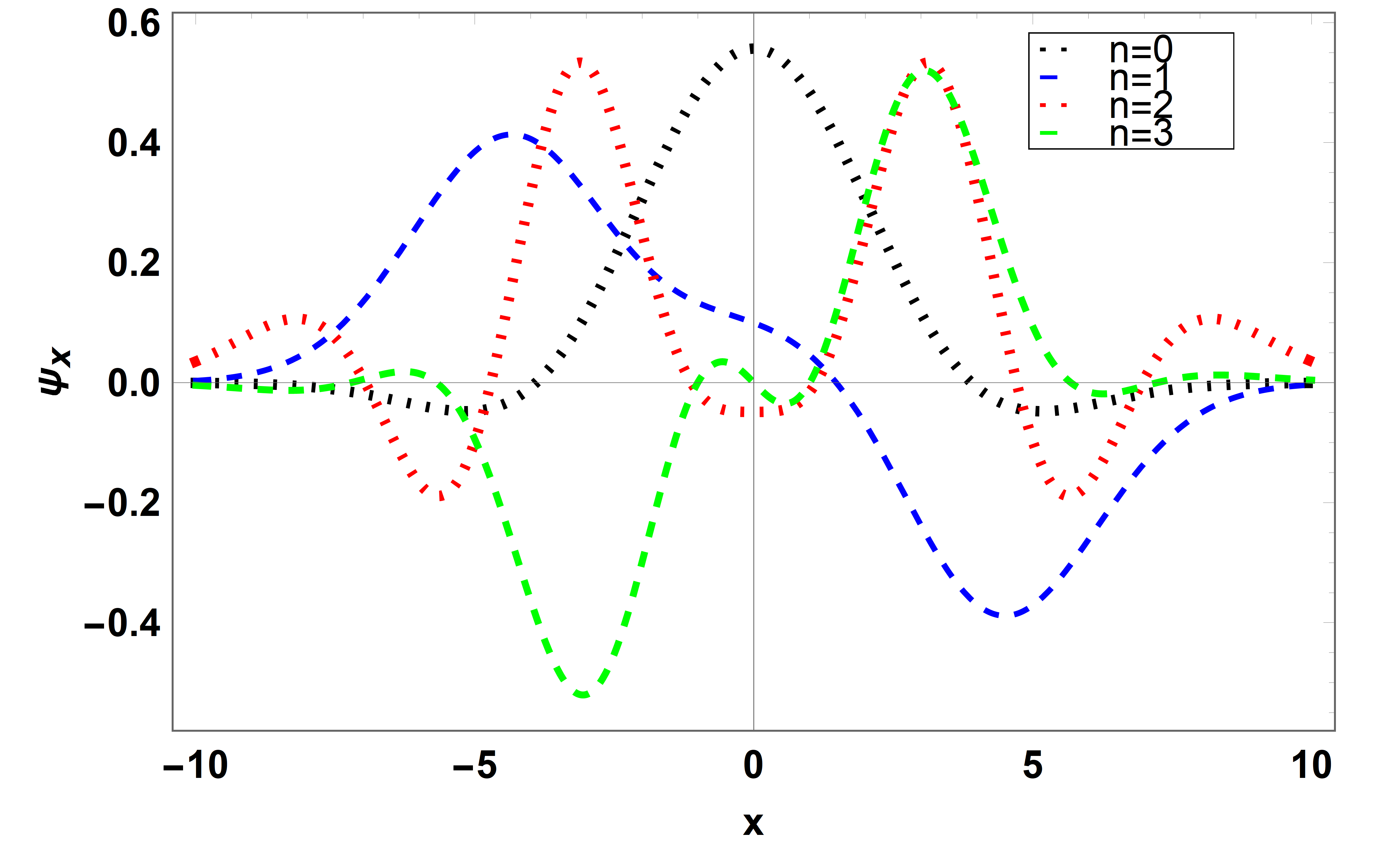}
  \caption{Position space}
  \label{fig:image112}
\end{subfigure}
\begin{subfigure}{0.5\textwidth}
  \centering
  \includegraphics[width=\linewidth]{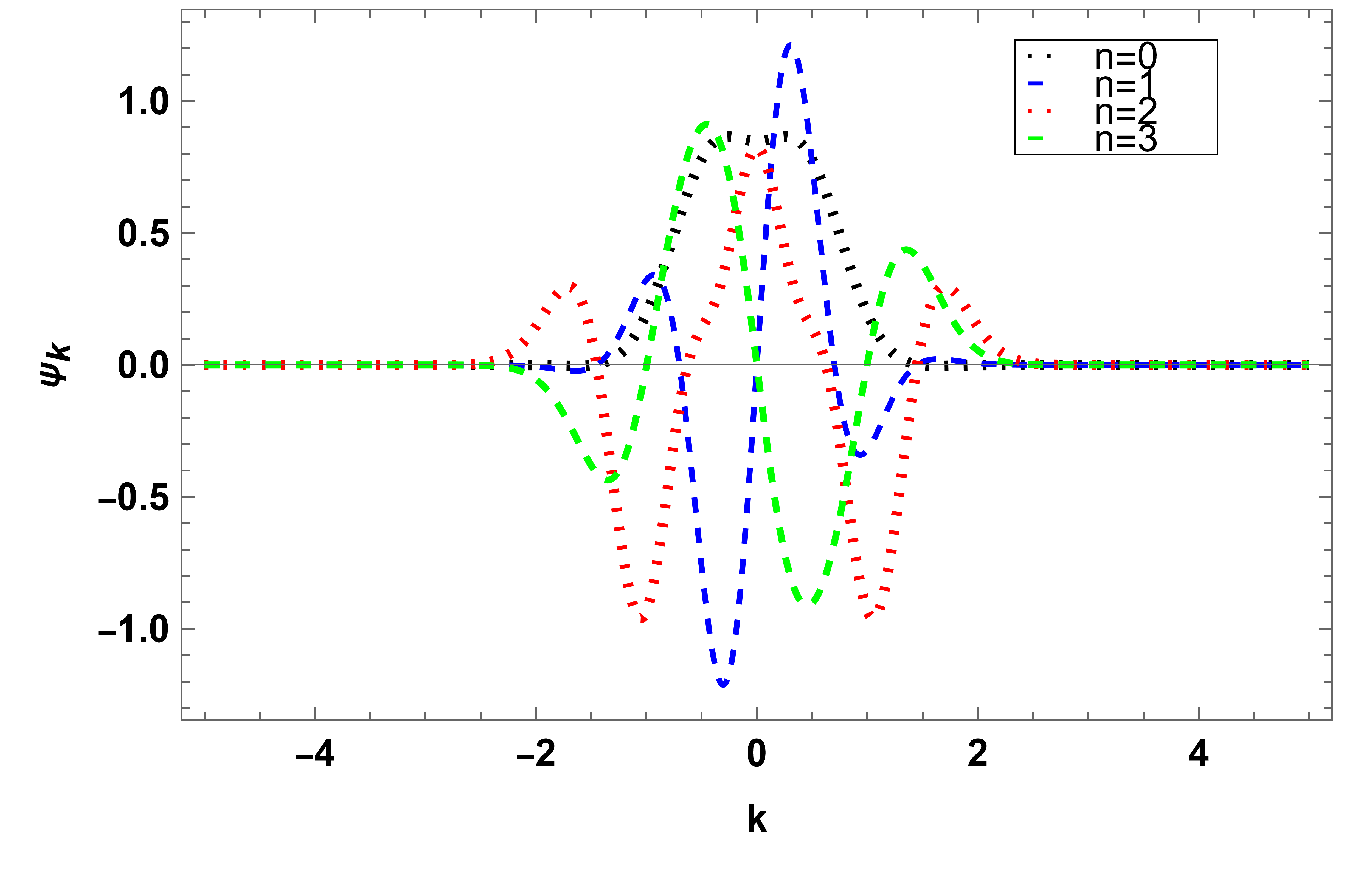}
  \caption{Momentum space}
   \label{fig:image27}
\end{subfigure}
 \caption{Wave functions of Symmetric well for various values of $n$.  }
 \label{fig:wavef2}
\end{figure}

\begin{figure}[H]
\begin{subfigure}{0.5\textwidth}
  \centering
  \includegraphics[width=\linewidth]{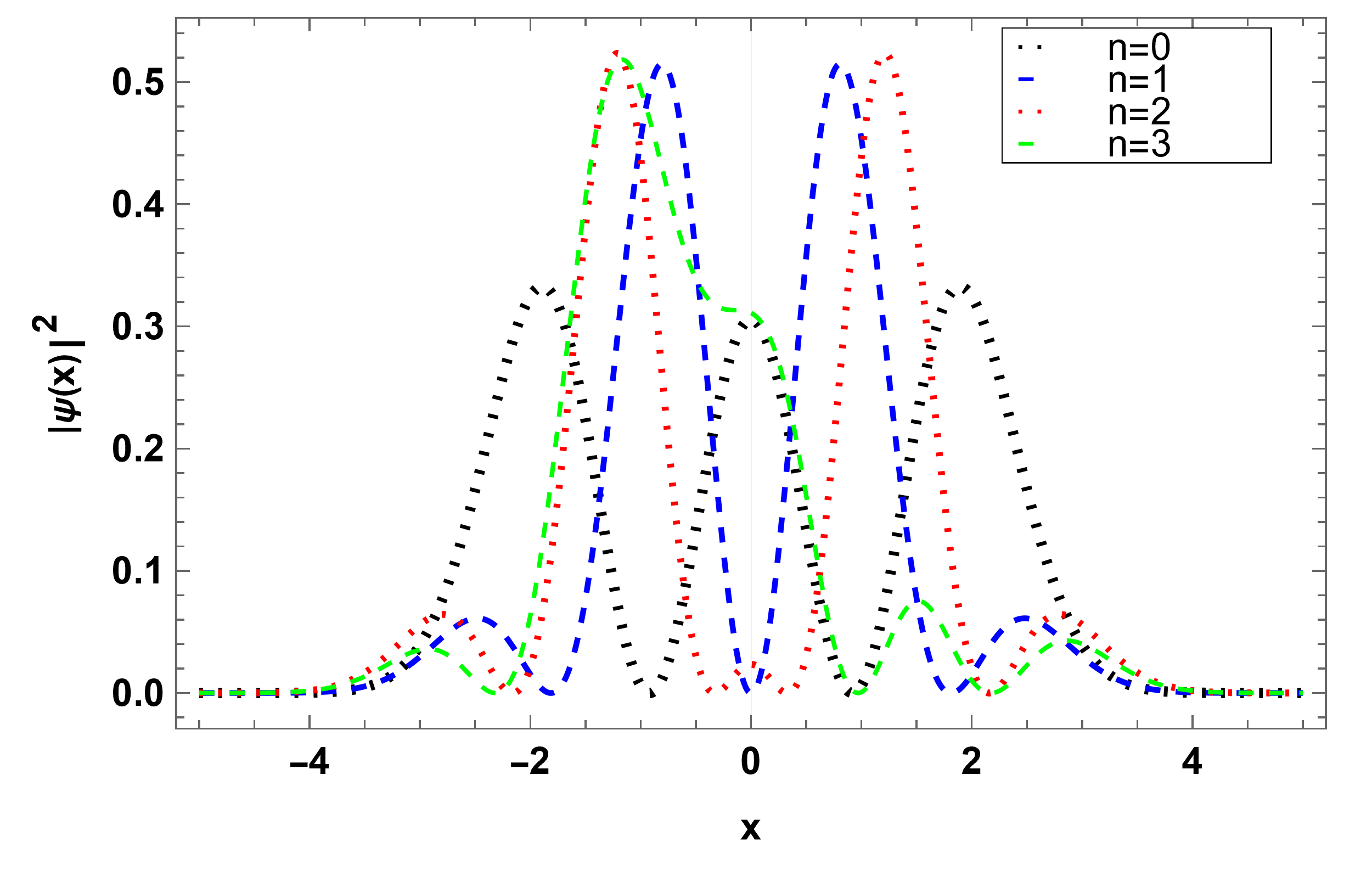}
  \caption{Position space}
  \label{fig:image1111}
\end{subfigure}
\begin{subfigure}{0.5\textwidth}
  \centering
  \includegraphics[width=\linewidth]{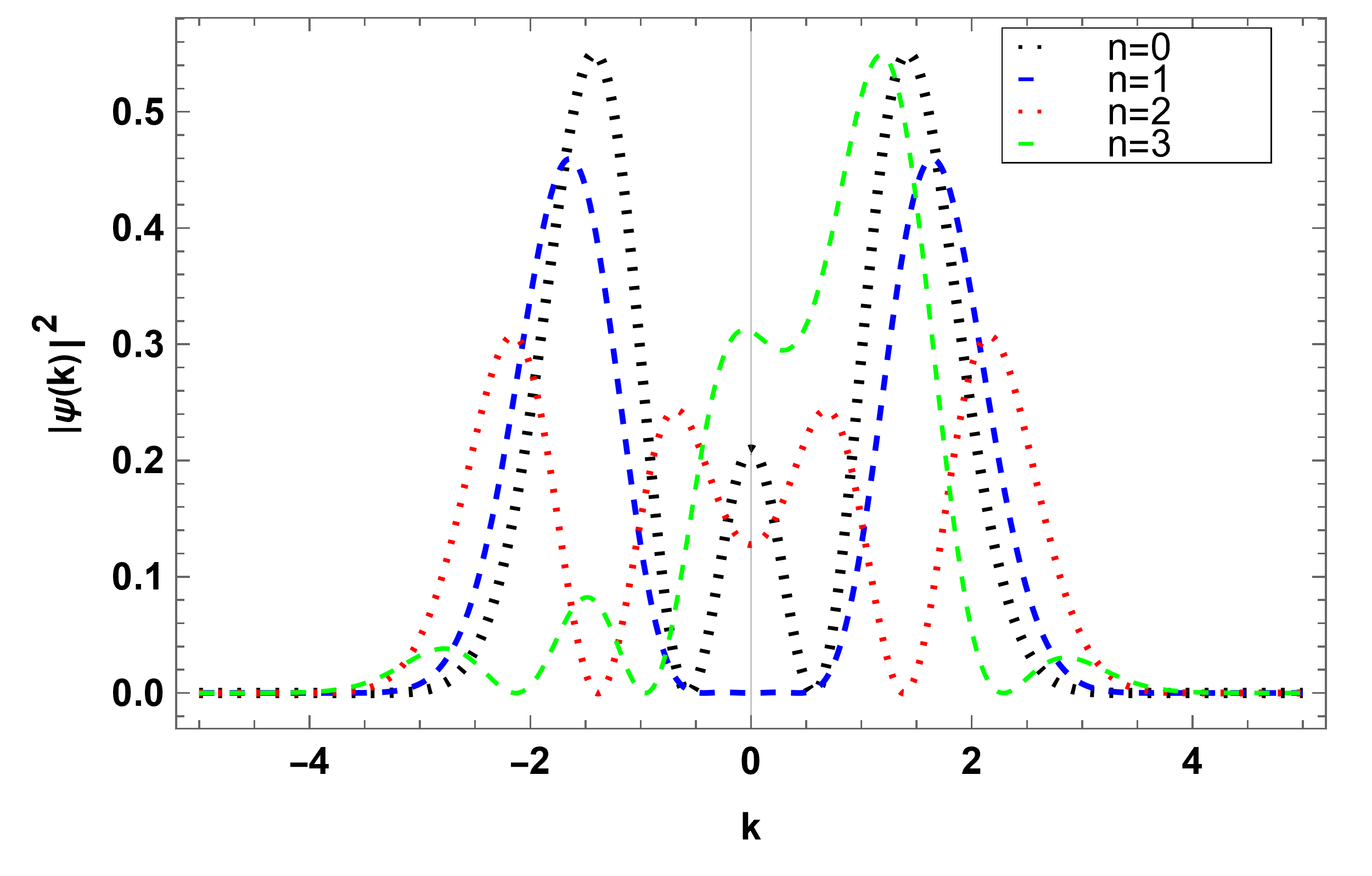}
  \caption{Momentum space}
   \label{fig:image261}
\end{subfigure}
 \caption{Probability densities for Quartic potential subjected to various values of $n$.  }
 \label{fig:image2main1}
\end{figure}

\begin{figure}[H]
\begin{subfigure}{0.5\textwidth}
  \centering
  \includegraphics[width=\linewidth]{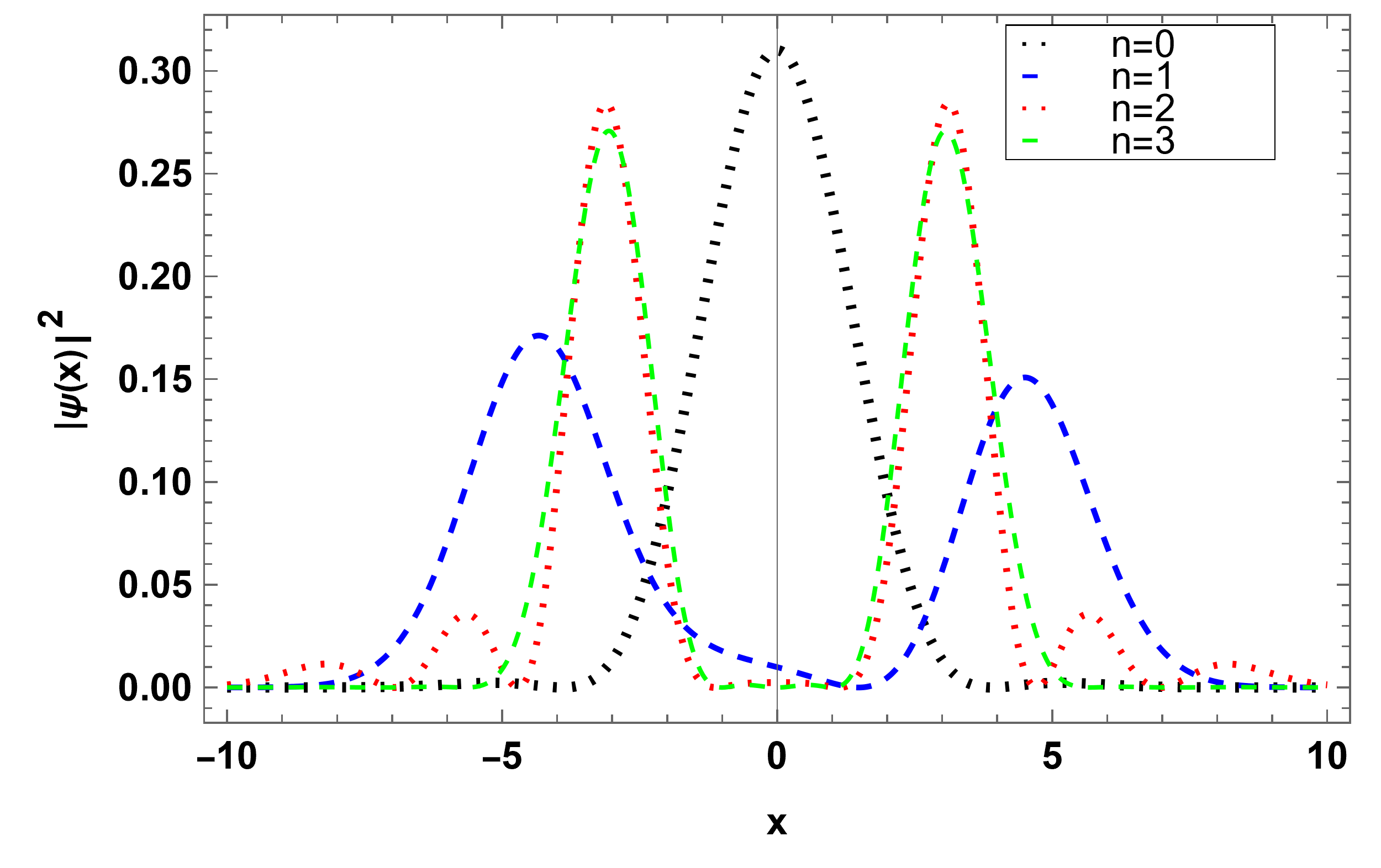}
  \caption{Position space}
  \label{fig:image1112}
\end{subfigure}
\begin{subfigure}{0.5\textwidth}
  \centering
  \includegraphics[width=\linewidth]{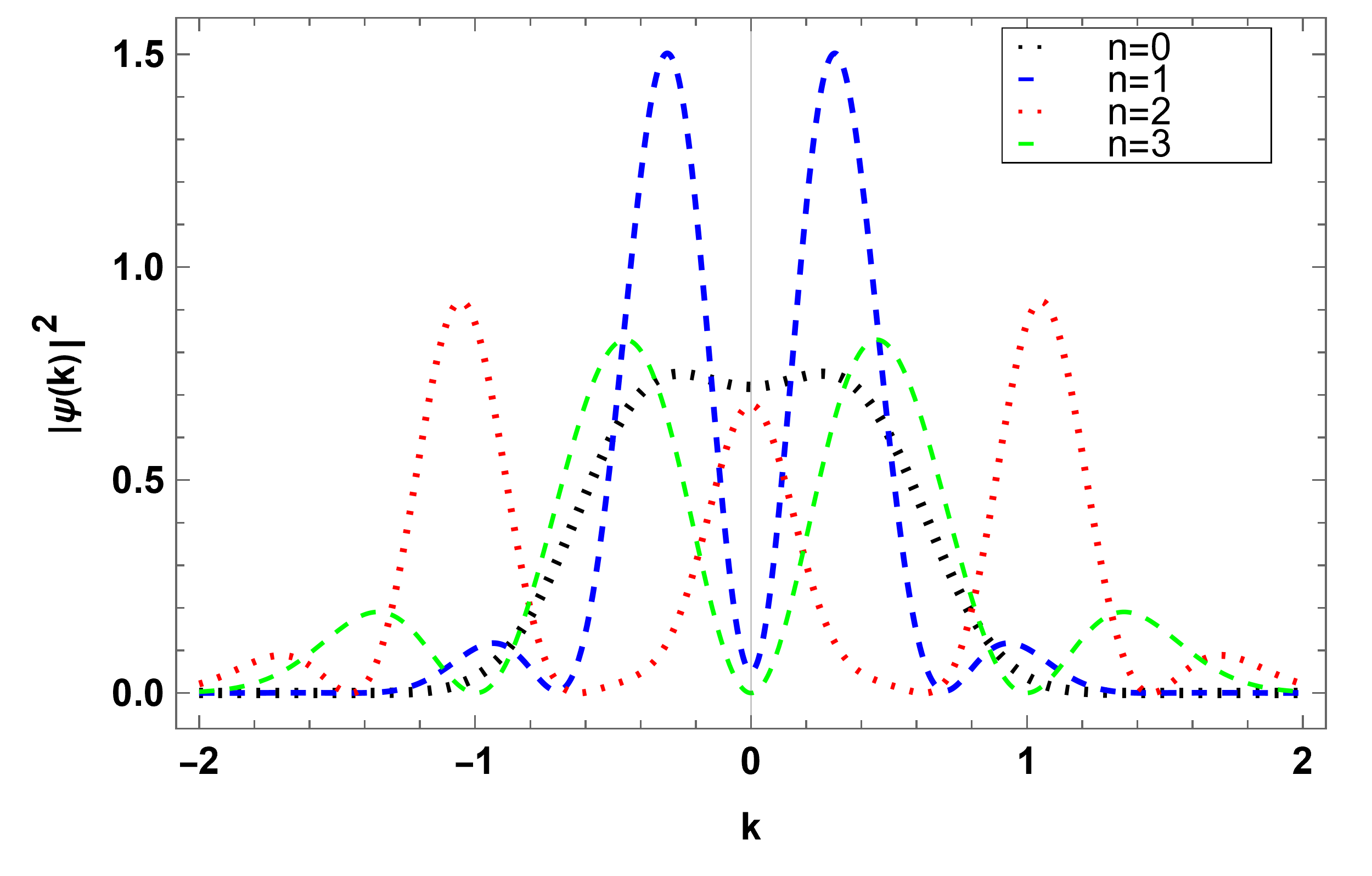}
  \caption{Momentum space}
   \label{fig:image262}
\end{subfigure}
 \caption{Probability densities for Symmetric well subjected to various values of $n$.  }
 \label{fig:image2main23}
\end{figure}

\section{Information measures} \label{sec:information}
The concept of entropy is interpreted in various ways across different fields. In thermodynamics, entropy serves as a measure of a system's irreversibility. Meanwhile, within the domain of information theory, Claude Shannon proposed the idea of Shannon entropy in his publication in 1948, \textquotedblleft A Mathematical Theory of Communication \textquotedblright \cite{heisenberg}. This metric quantifies the average information output from a stochastic source. Essentially, a greater Shannon entropy value signifies that each subsequent outcome of the process carries a higher amount of information. Formally, for a probability distribution $p_i$, Shannon entropy is expressed as \cite{heisenberg}:
\begin{equation} \label{defnshannon}
    S= -\sum p_{i}\text{ln}(p_{i}).
\end{equation}
Shannon entropy was originally formulated as an independent concept, without relying on any prior assumptions, specifically to quantify uncertainty. It resembles Gibbs entropy and is sometimes known as \textquotedblleft Boltzmann-Gibbs-Shannon  \textquotedblright entropy \cite{imp3}. This measure captures the uncertainty in a probability distribution linked to an information source. When considering a probability distribution in position space, the Shannon entropy in the $n$th state is given by \cite{qiang2010proper}:
\begin{equation}\label{nxshannon}
    S_{x} = -\int_{-\infty}^{\infty} dx \bigg[(\psi_{n}(x))^{2} \text{ln} (\psi_{n}(x))^{2}\bigg].
\end{equation}
In the momentum (reciprocal) space, the Shannon entropy corresponding to the $n$th state is given by: 
\begin{equation}\label{pnshannon}
      S_{k} = -\int_{-\infty}^{\infty} dk \bigg[(\phi_{n}(k))^{2} \text{ln} (\phi_{n}(k))^{2}\bigg],
\end{equation}
where,
\begin{equation} \label{fourier}
\phi_{n}(k) = \frac{1}{\sqrt{2\pi}}\int_{-\infty}^{\infty} e^{-ikx} \psi_{n}(x) dx.  
\end{equation}

Although originating from different disciplines, the uncertainty principle and Shannon entropy both address the theme of uncertainty and the inherent limits of what can be known. In quantum mechanics, the uncertainty principle defines the fundamental boundaries on how accurately we can measure paired properties, such as position and momentum or energy and time, simultaneously. This principle arises from the dual characteristics of particles and waves and the inherently probabilistic behavior of quantum systems. On the other hand, Shannon entropy measures the amount of information in a probabilistic system, highlighting how unpredictable the outcomes are. Despite their distinct backgrounds, both concepts highlight the inherent limitations on measurement and knowledge.

Shannon entropy measures the degree of unpredictability or dispersion within a probability distribution, reflecting the average amount of information required to describe or predict the outcomes of a random variable. A higher entropy value signifies greater unpredictability, while a lower value indicates less uncertainty and more accurate forecasting. Even though these concepts are not directly related, they both highlight the inherent limitations in our capacity to fully comprehend or forecast specific aspects of physical systems. The uncertainty principle imposes limits on our capacity to measure paired variables simultaneously in quantum mechanics, while Shannon entropy assesses the level of uncertainty or information content in random variables within a probabilistic framework. This connection is briefly explored in \cite{unentropy1, unentropy2} and illustrated as follows:
\begin{equation} \label{uncertain}
    S_{x}+S_{k} \geq D(1+\text{ln} \pi).
\end{equation}
Here, $D$ represents the dimension of a given system. Determining Shannon entropy using analytical techniques can be quite difficult due to the complexity introduced by the logarithmic terms within the integrals, particularly when factoring in the system's dimension. Consequently, explicit results have been obtained only for certain low-energy states of specific systems, such as the Harmonic oscillator \cite{harmonic}, the Pöschl–Teller potential \cite{poschl1}, the Morse potential \cite{morse1}, and highly excited states of the Coulomb potential \cite{coulomb}. Entropy is a fundamental principle that supports the second law of thermodynamics \cite{entropy}. This principle suggests that in a closed system, the overall entropy tends to increase as time goes on, eventually reaching its maximum value as time approaches infinity. In its classical interpretation, entropy is defined for systems at equilibrium, as originally formulated by Clausius and others (refer to \cite{imp4}), and it is illustrated as:
\begin{equation} \label{entropy}
    dS = \frac{\delta Q}{T}.
\end{equation}
As noted earlier, information theory employs entropy to measure the degree of disorder in a system \cite{unentropy1}. Thus, in a thermodynamic context, entropy increases with the number of potential microstates accessible to the system. Additionally, information theory introduces the concept of Fisher information ($F$) \cite{fisher1}. Fisher information measures the quantity of information that a random variable provides about an unknown parameter on which the probability distribution is based. The mathematical expression for Fisher information is: 
\begin{equation} \label{fisherintro}
    F = \sum_{i=1}^{m} \frac{1}{p_{i}}\bigg[\frac{dp_{i}}{di}\bigg]^{2}.
\end{equation}
In this context, $p_i$ represents the likelihood of the system being in microstate $i$. Within a thermodynamic framework \cite{thermo}, as the number of possible microstates increases, the system’s entropy also increases. This rise in entropy signifies a greater degree of randomness or disorder in the system. For example, if the system is confined to a single microstate, the probability density function will exhibit a pronounced peak at that microstate, indicating a high level of Fisher information. In contrast, if the system is spread across many microstates with nearly uniform probabilities, the probability density function flattens out, resulting in a near-zero slope and thus lower Fisher information.

Quantum metrology is a field focused on using quantum mechanics to achieve extremely precise measurements, going beyond the limits of classical methods. A key concept in this area is the quantum Fisher information, which helps define the ultimate precision achievable and plays a crucial role in analyzing measurement strategies \cite{liu2020quantum}. It is also useful for studying quantum entanglement, which is essential for improving measurement accuracy, especially in systems involving multiple parameters. Entangled states like Greenberger-Horne-Zeilinger (GHZ) \cite{greenberger1989going} and spin squeezed states \cite{ma2011quantum} allow for higher precision by taking advantage of quantum correlations. These states are crucial in applications like magnetometry and interferometry, where precise measurements are critical \cite{toth2014quantum}.

So, Fisher's information helps us to understand how organized a system is by identifying situations where the system is reliably found in one or a few microstates with high probability and help understand and detect the role of entanglement in enhancing performance in these tasks. It is crucial to understand that, although equations (\ref{defnshannon}) and (\ref{fisherintro}) may seem similar, Shannon entropy and Fisher information assess different aspects of a system. Shannon entropy depends solely on the probability density function itself, while Fisher information is affected by the derivative of this function, which reflects its slope. Therefore, Shannon entropy can be viewed as a global characteristic of the system, In contrast, Fisher information provides a more specific measure related to the probability density function. Therefore, Fisher's information and Shannon's entropy are not directly comparable. Fisher information for an observable $\zeta$ is defined as:
\begin{equation}\label{fisher}
    F_{\zeta} = \int_{-\infty}^{\infty} d\zeta \bigg[\rho(\zeta)\bigg(\frac{d}{d\zeta}\text{ln}(\rho(\zeta))\bigg)^{2}\bigg] > 0.
\end{equation}
We determine Fisher information for an observable $\zeta$ with a probability density expressed as $|\psi(\zeta,t)|^{2}$ is given by
  $$  F_{\zeta} = \int_{-\infty}^{\infty} d\zeta \bigg[(\psi(\zeta, t))^{2} \bigg(\frac{d}{d\zeta}\text{ln} (\psi(\zeta,t))^{2}\bigg)^{2}\bigg] > 0. $$
  
  For one-dimensional stationary quantum systems, like solitonic systems, the probability density in the position space $\rho(\zeta)$ is given by  $|\psi(\zeta,t)|^{2}$ which can be approximated as $|\psi(\zeta)|^{2} $. This approximation arises from treating $\psi(\zeta,t)$ as approximately equal to  $\psi(\zeta)$.
  
  \begin{equation}
      F_{\zeta} = \int_{-\infty}^{\infty} d\zeta \bigg[(\psi(\zeta))^{2} \bigg(\frac{d}{d\zeta}\text{ln} (\psi(\zeta))^{2}\bigg)^{2}\bigg].
  \end{equation}
  
In a similar manner, we calculate Fisher information for the same soliton in momentum space.
\begin{equation} \label{fishermomentum}
F_{k} = \int_{-\infty}^{\infty} dk \bigg[(\phi(k, t))^{2} \bigg(\frac{d}{dk}\text{ln} (\phi(k,t))^{2}\bigg)^{2}\bigg],
\end{equation}  
By applying the same approximation as previously, we obtain the Fisher information:
\begin{equation}
    F_{k} = \int_{-\infty}^{\infty} dk \bigg[(\phi(k))^{2} \bigg(\frac{d}{dk}\text{ln} (\phi(k))^{2}\bigg)^{2}\bigg].
\end{equation}
  where, $\phi(k) = \frac{1}{\sqrt{2\pi}}\int_{-\infty}^{\infty} e^{-ik\zeta} \psi(\zeta) d\zeta $. In terms of Fisher information, uncertainty refers to the degree of inaccuracy or lack of complete understanding regarding the true value of a parameter. It reflects the variation or dispersion in estimates derived from various samples or observations. This concept assesses how uncertain or unpredictable parameter estimation is based on empirical data. Greater Fisher information indicates reduced uncertainty and more precise estimates. This is demonstrated by:

\begin{equation}\label{eq:infouncertain}
     F_{\zeta}F_{k} \geq 4\hbar^{2}.
\end{equation}
where $\hbar$ is reduced Planck's constant.
\subsection{Shannon entropy} \label{sec:shannon}

As stated by Bonn \cite{102}, From a statistical perspective, the likelihood of finding the particle in a specific state $\psi(x,t)$ within the spatial interval from $x$ to $x + dx$ is approximated by $|\psi(x,t)|^{2} \approx |\psi(x)|^{2}$. The Shannon entropy in position space ($S_x$) is given by:
\begin{equation}\label{eq:150}
    S_{x} = -\int_{-\infty}^{\infty} |\psi(x)|^{2} \text{ln}[|\psi(x)|^{2}] dx.
\end{equation}  

\begin{figure}[H]
    \centering
    \begin{subfigure}[t]{0.45\textwidth}
        \centering
        \includegraphics[width=\textwidth]{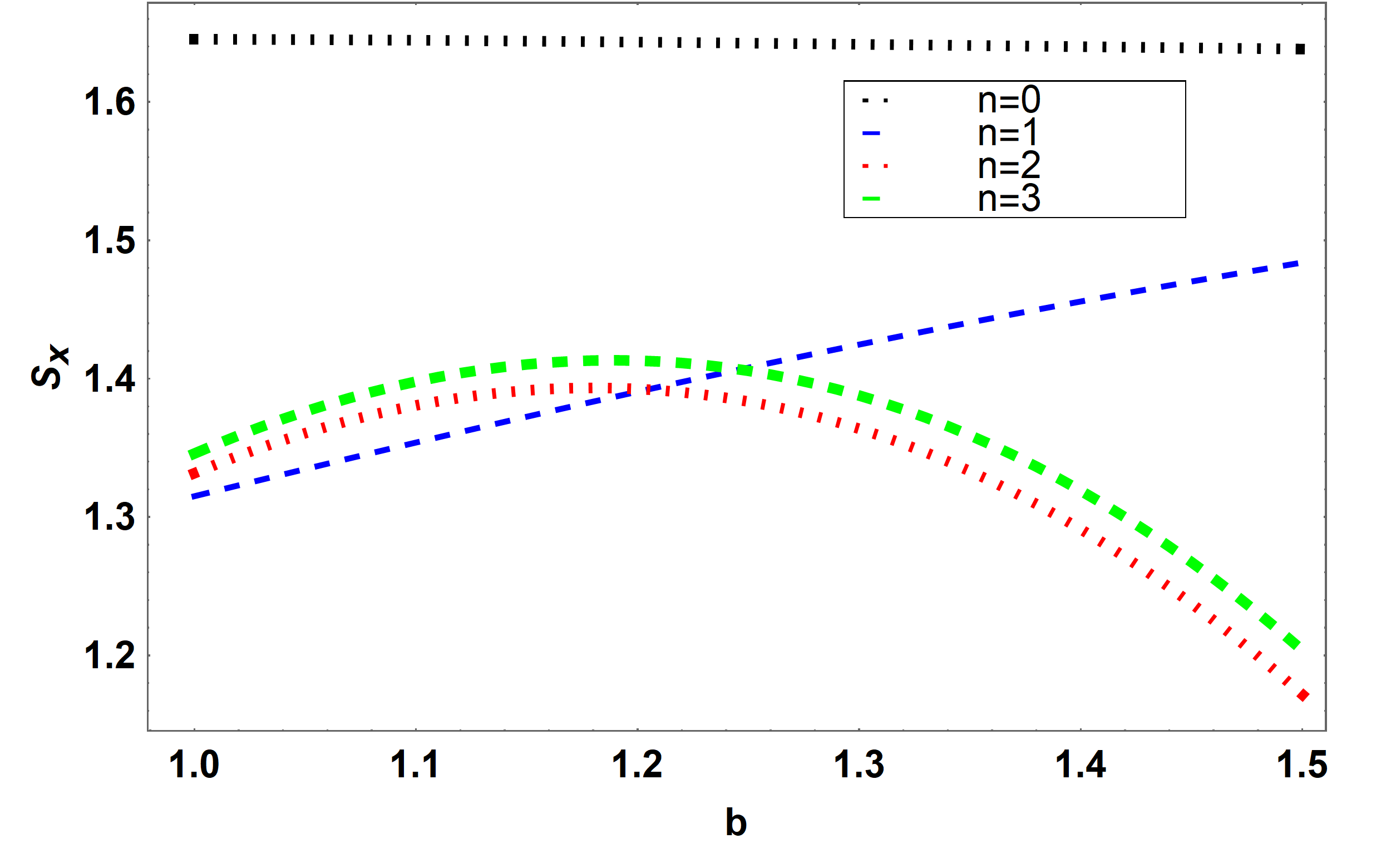} 
        \caption{Position space}
         \label{fig:image3}
    \end{subfigure}
    \hfill
    \begin{subfigure}[t]{0.45\textwidth}
        \centering
        \includegraphics[width=\textwidth]{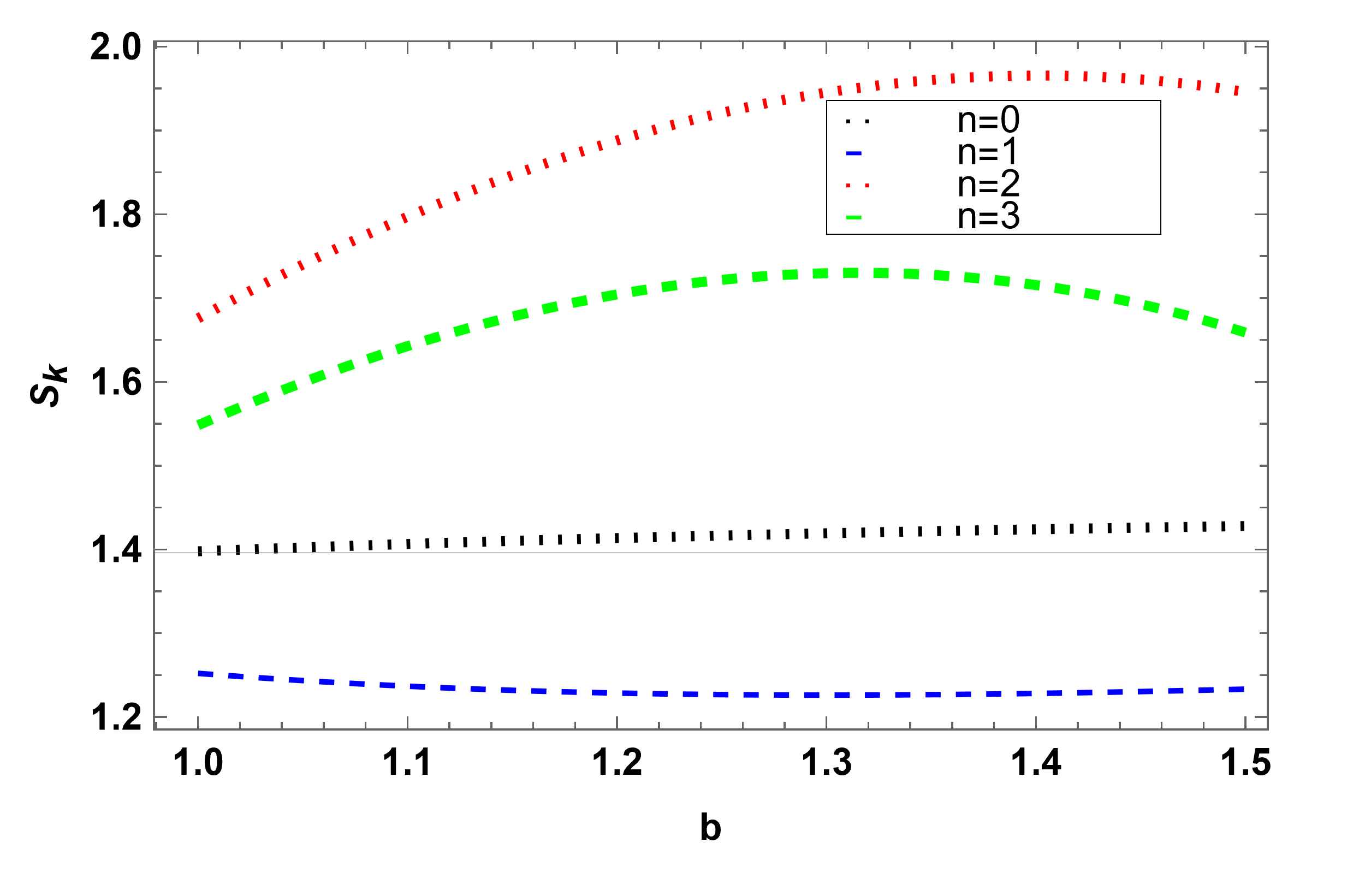} 
        \caption{Momentum space}
         \label{fig:image4}
    \end{subfigure}

    \vspace{1em} 
    \begin{subfigure}[t]{0.5\textwidth}
        \centering
        \includegraphics[width=\textwidth]{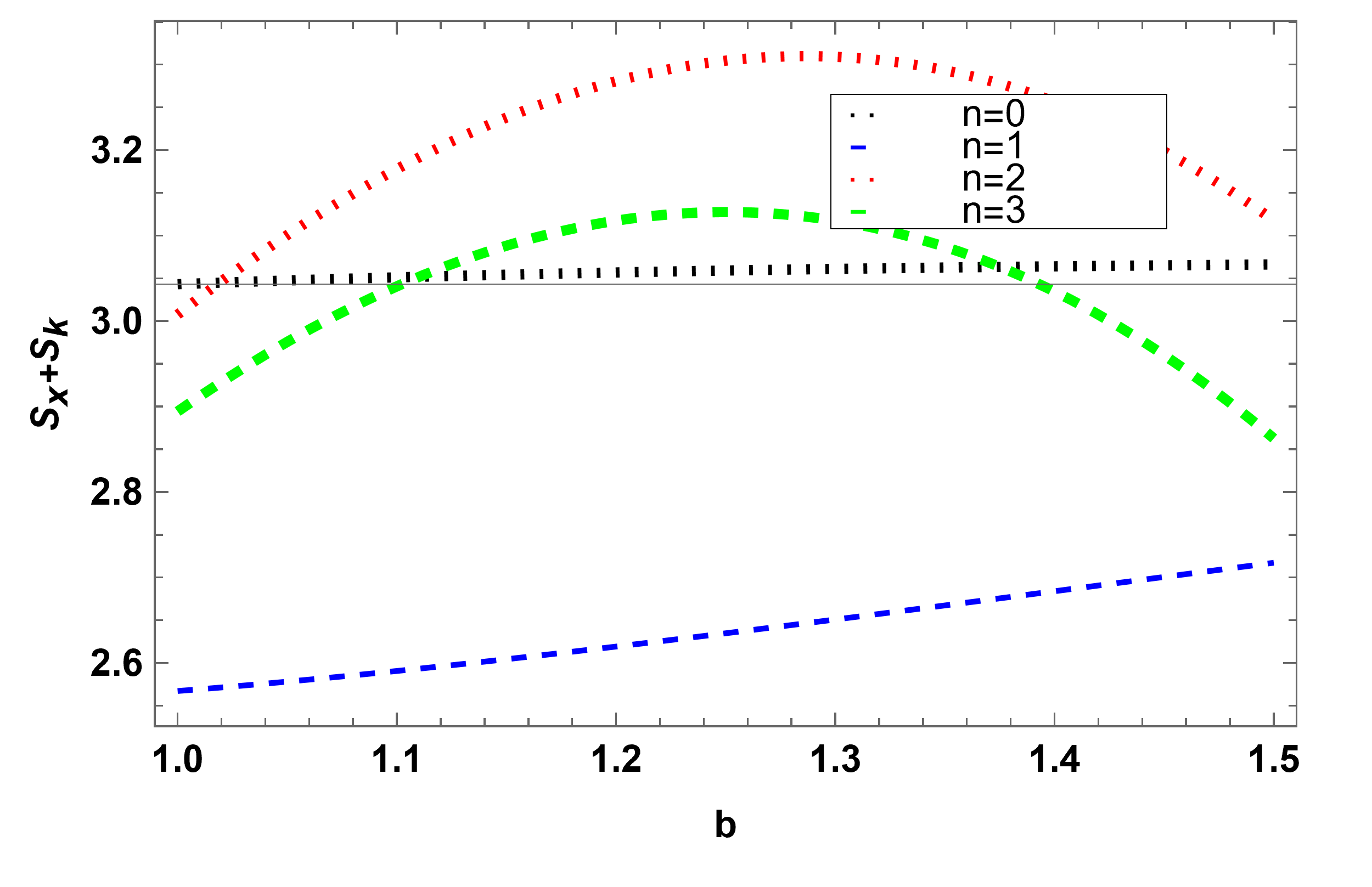} 
        \caption{Combined value of $S_x+S_k$}
         \label{fig:image5}
    \end{subfigure}

    \caption{Plots of Shannon entropy for quartic potential for different values of $b$ and $n$.}
     \label{fig:image2main2}
\end{figure}

\begin{figure}[H]
    \centering
    \begin{subfigure}[t]{0.45\textwidth}
        \centering
        \includegraphics[width=\textwidth]{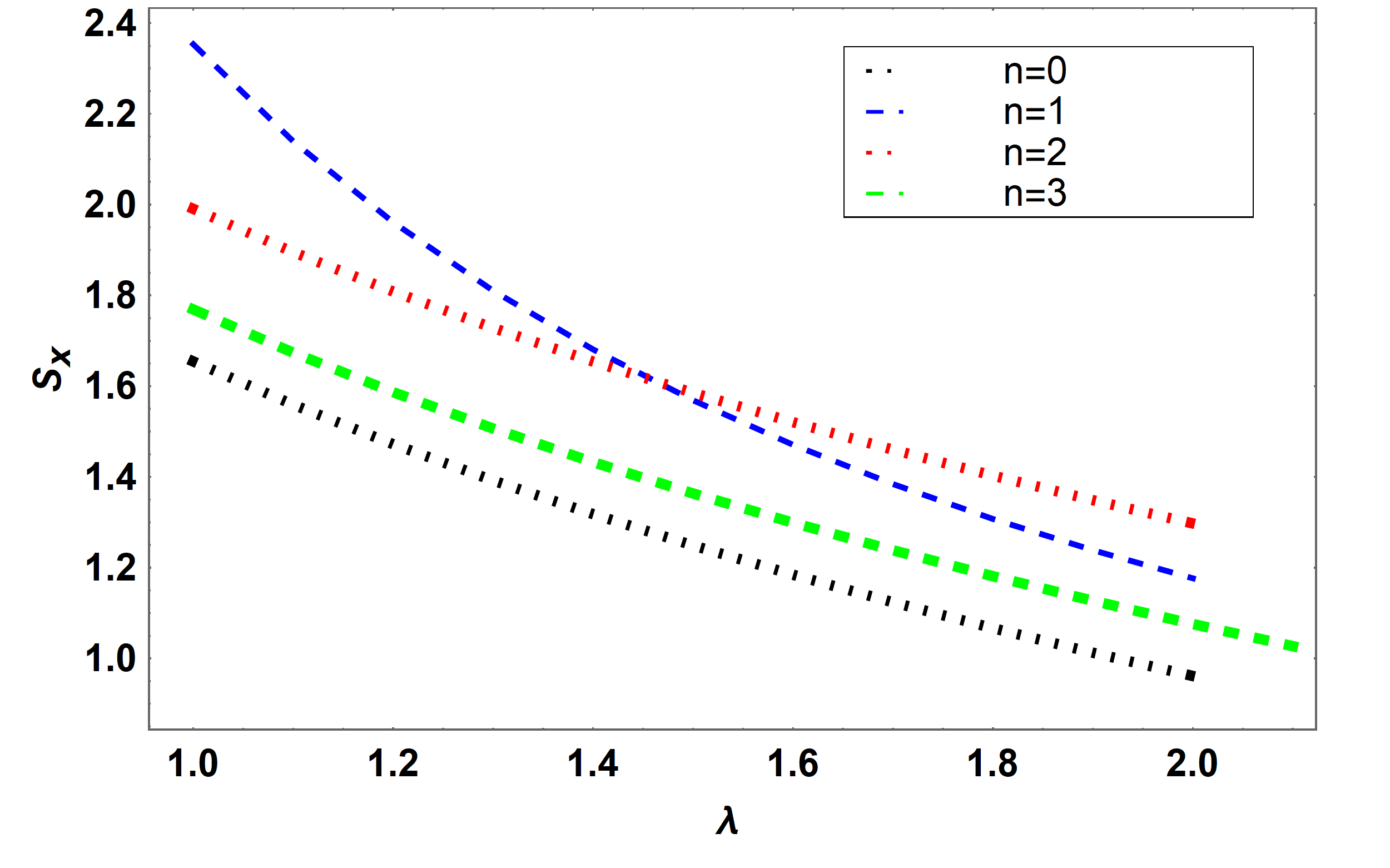} 
        \caption{Position space}
         \label{fig:image6}
    \end{subfigure}
    \hfill
    \begin{subfigure}[t]{0.45\textwidth}
        \centering
        \includegraphics[width=\textwidth]{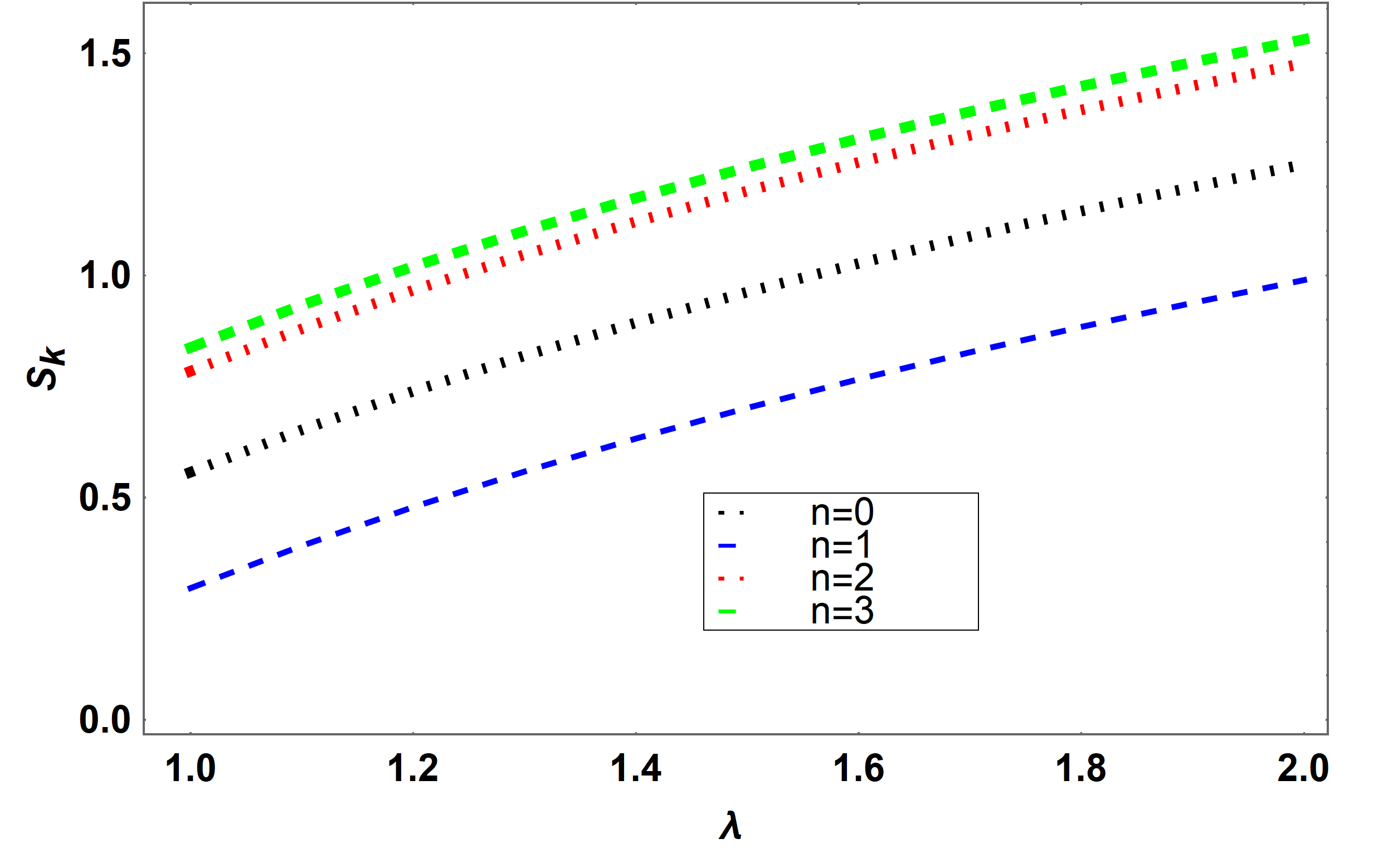} 
        \caption{Momentum space}
         \label{fig:image7}
    \end{subfigure}

    \vspace{1em} 
    \begin{subfigure}[t]{0.5\textwidth}
        \centering
        \includegraphics[width=\textwidth]{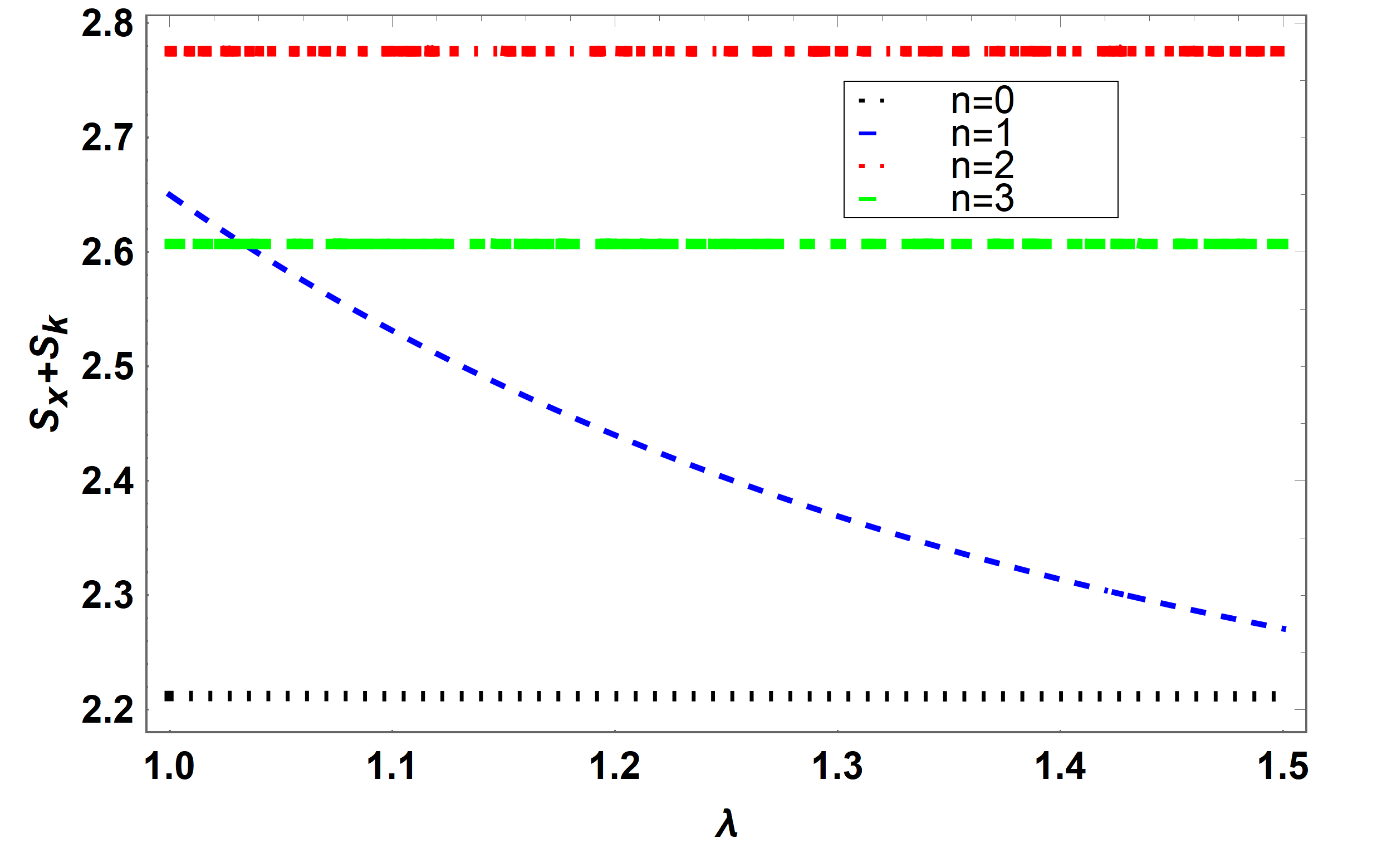} 
        \caption{Combined value of $S_x+S_k$}
         \label{fig:image8}
    \end{subfigure}

    \caption{Plots of Shannon entropy for symmetric well for different values of $b$ and $n$.}
     \label{fig:image2main3}
\end{figure}

 \noindent In momentum space, it is represented by $S_k$
\begin{equation} \label{eq:151}
    S_{k} = -\int_{-\infty}^{\infty} |\phi(k)|^{2} \text{ln}[|\phi(k)|^{2}] dk, 
\end{equation}
\begin{equation} 
    S_{x} + S_{k} \geq D(1+\text{ln}\pi). \nonumber
\end{equation}
The computation of Shannon entropy was carried out by analyzing the eigenfunctions in both the position and momentum space, using the definitions provided in equations \eqref{eq:150} and \eqref{eq:151}. The results for Shannon entropy in the context of quartic and symmetric wells are detailed in tables \ref{shannontable1} and \ref{shannontable2}, respectively. Additionally, the plots illustrating Shannon entropy for quartic and symmetric wells as a function of the parameter 
$b$ across various states are shown in figures  \ref{fig:image2main2} and \ref{fig:image2main3}.
\begin{center}
\begin{table}[ht]
    \begin{tabular}{|p{2cm}<{\centering}|p{3cm}<{\centering}|p{2cm}<{\centering}|p{2cm}<{\centering}|p{4cm}<{\centering}|p{3cm}<{\centering}|}
        \hline
        \textbf{n} & \textbf{b} & \textbf{$S_x$} & \textbf{$S_k$} & \textbf{$S_x+S_k$} & \textbf{$1+\ln \pi$} \\ \hline
        
        \multirow{3}{*}{0} & 1    & 1.64536 & 1.39774 & 3.0431  & 2.1447 \\ 
                             & 1.3    & 1.64169 & 1.41926   & 3.06095     & 2.1447 \\ 
                             & 1.5 & 1.63824 & 1.42792 & 3.06616   & 2.1447 \\ \hline
                             
        \multirow{3}{*}{1} & 1    & 1.3151 & 1.25211   & 2.56721    & 2.1447 \\ 
                             & 1.3      & 1.42467 & 1.22618 & 2.65085   & 2.1447 \\ 
                             & 1.5   & 1.48395 & 1.23328   &2.71723 & 2.1447 \\ \hline
                             
        \multirow{3}{*}{2} & 1   & 1.33173 & 1.67577 & 3.0075          & 2.1447 \\ 
                             & 1.3 & 1.36383 & 1.94526 & 3.30909      & 2.1447 \\ 
                             & 1.5  & 1.17179 & 1.94589   & 3.11768     & 2.1447 \\ \hline
                             
        \multirow{3}{*}{3} & 1  & 1.34594 & 1.5482 & 2.8941   & 2.1447 \\ 
                             & 1.3   & 1.38785 & 1.72967   & 3.1175 & 2.1447 \\ 
                             & 1.5   & 1.20378 & 1.65886  &2.8625  & 2.1447 \\ \hline

    \end{tabular}
    \caption{Numerical values of Shannon entropy for the quartic potential}
    \label{shannontable1}
    \end{table}
\end{center}

\begin{center}
\begin{table}[ht]
    \begin{tabular}{|p{2cm}<{\centering}|p{3cm}<{\centering}|p{2cm}<{\centering}|p{2cm}<{\centering}|p{4cm}<{\centering}|p{3cm}<{\centering}|}
        \hline
        \textbf{n} & \textbf{$\lambda$} & \textbf{$S_x$} & \textbf{$S_k$} & \textbf{$S_x+S_k$} & \textbf{$1+\ln \pi$} \\ \hline
        
        \multirow{3}{*}{0} & 1    & 1.65452 & 0.557508 & 2.21202  & 2.1447 \\ 
                             & 1.3    & 1.39217 & 0.819666   & 2.2118     & 2.1447 \\ 
                             & 1.5 & 1.24908 & 0.962654 & 2.21173   & 2.1447 \\ \hline
                             
        \multirow{3}{*}{1} & 1    & 2.35342 & 0.296498   & 2.6499    & 2.1447 \\ 
                             & 1.3      & 1.81032 & 0.558854 & 2.3691   & 2.1447 \\ 
                             & 1.5   & 1.56895 & 0.701946    & 2.2708 & 2.1447 \\ \hline
                             
        \multirow{3}{*}{2} & 1   & 1.99074 & 0.784698 & 2.77543          & 2.1447 \\ 
                             & 1.3 & 1.72838 & 1.04706 & 2.77544      & 2.1447 \\ 
                             & 1.5  & 1.58528 & 1.19016  & 2.77544     & 2.1447 \\ \hline
                             
        \multirow{3}{*}{3} & 1  & 1.76912 & 0.838033 & 2.60715   & 2.1447 \\ 
                             & 1.3   & 1.50675 & 1.1004   & 2.60715 & 2.1447 \\ 
                             & 1.5   & 1.36365 & 1.2435   &2.60715 & 2.1447 \\ \hline

    \end{tabular}
    \caption{Numerical values of Shannon entropy for the symmetric well}
    \label{shannontable2}
    \end{table}
\end{center}

\subsection{Fisher information measures} \label{sec:fisher} 
 For an observable $x$, Fisher information ($F_{x}$) is expressed as
\begin{equation}
    F_{x} = \int_{-\infty}^{\infty} |\psi(x)|^{2}\bigg[\frac{d}{dx}\text{ln}|\psi(x)|^{2}\bigg]^{2} dx  > 0. \nonumber
\end{equation}

In momentum space, the Fisher information for an observable $k$, represented as $F_k$ is specified by:
\begin{equation}
    F_{k} = \int_{-\infty}^{\infty} |\phi(k)|^{2}\bigg[\frac{d}{dk}\text{ln}|\phi(k)|^{2}\bigg]^{2} dk > 0. \nonumber
\end{equation}

It is also important to note that the standard deviations for position and momentum measurements are defined as follows:

\begin{equation} \label{eq:170}
    \sigma^{2}_{x} = \langle x^{2}\rangle-\langle x\rangle ^{2}, 
\end{equation}
\begin{equation}\label{eq:171}
     \sigma^{2}_{k} = \langle k^{2}\rangle-\langle k\rangle ^{2} .
\end{equation}

where, $\langle x\rangle, \langle k\rangle, \langle x^{2}\rangle, \langle k^{2}\rangle$ denote the expected values of $x, k, x^{2}, k^{2}$ respectively. By applying the Fisher information definitions provided in equations \eqref{fisher} and \eqref{fishermomentum}, along with the standard deviations given in equations \eqref{eq:170} and \eqref{eq:171}, we present the numerical results for Fisher information as shown in the table below:

\begin{center}
\begin{table}[ht]
    \begin{tabular}{|p{1cm}<{\centering}|p{3cm}<{\centering}|p{2cm}<{\centering}|p{2cm}<{\centering}|p{2cm}<{\centering}|p{2cm}<{\centering}|p{2cm}<{\centering}|}
        \hline
        \textbf{n} & \textbf{b} & \textbf{$F_x$} & \textbf{$F_k$} & \textbf{$\sigma^{2}_{x}$} & \textbf{$\sigma^{2}_{k}$} & \textbf{$F_{x}\cdot F_{k}$}\\ \hline
        
        \multirow{3}{*}{0} & 1    & 8.25641 & 13.0727 & 3.2681  & 2.0641 & 107.933\\ 
                             & 1.3    & 8.10011 & 13.6546   & 3.4136     & 2.0250 & 110.6037 \\ 
                             & 1.5 & 8.0176 & 13.9005 & 3.4751   & 2.0044 & 111.4486 \\ \hline
                             
        \multirow{3}{*}{1} & 1    & 12.3848 & 5.99988   & 1.4999    & 3.0962 & 74.3073\\ 
                             & 1.3      & 13.1106 & 7.43391 & 1.8584   & 3.2776 & 97.4630\\ 
                             & 1.5   & 13.373 & 8.35914   & 2.0897 & 3.3432 & 111.7867 \\ \hline
                             
        \multirow{3}{*}{2} & 1   & 12.817 & 9.56661 & 2.3916         & 3.2042 & 122.6152 \\ 
                             & 1.3 & 21.6608 & 16.1676 & 4.0419      & 5.4152 & 350.2031\\ 
                             & 1.5  & 28.8383 & 21.5249   & 5.3812    & 4.7095 & 405.4952 \\ \hline
                             
        \multirow{3}{*}{3} & 1  & 6.03865 & 3.23455 & 0.8086   & 1.50966 & 19.5323\\ 
                             & 1.3   & 10.2053 & 5.46639   & 1.3665 & 2.5513 & 55.7861 \\ 
                             & 1.5   & 13.587 & 7.27774   & 1.8194  & 3.39675 & 98.8821\\ \hline

    \end{tabular}
    \caption{Numerical values of the Fisher's Information measure for quartic potential}
    \label{fishertable1}
    \end{table}
\end{center}

\begin{figure}[H]
    \centering
    \begin{subfigure}[t]{0.45\textwidth}
        \centering
        \includegraphics[width=\textwidth]{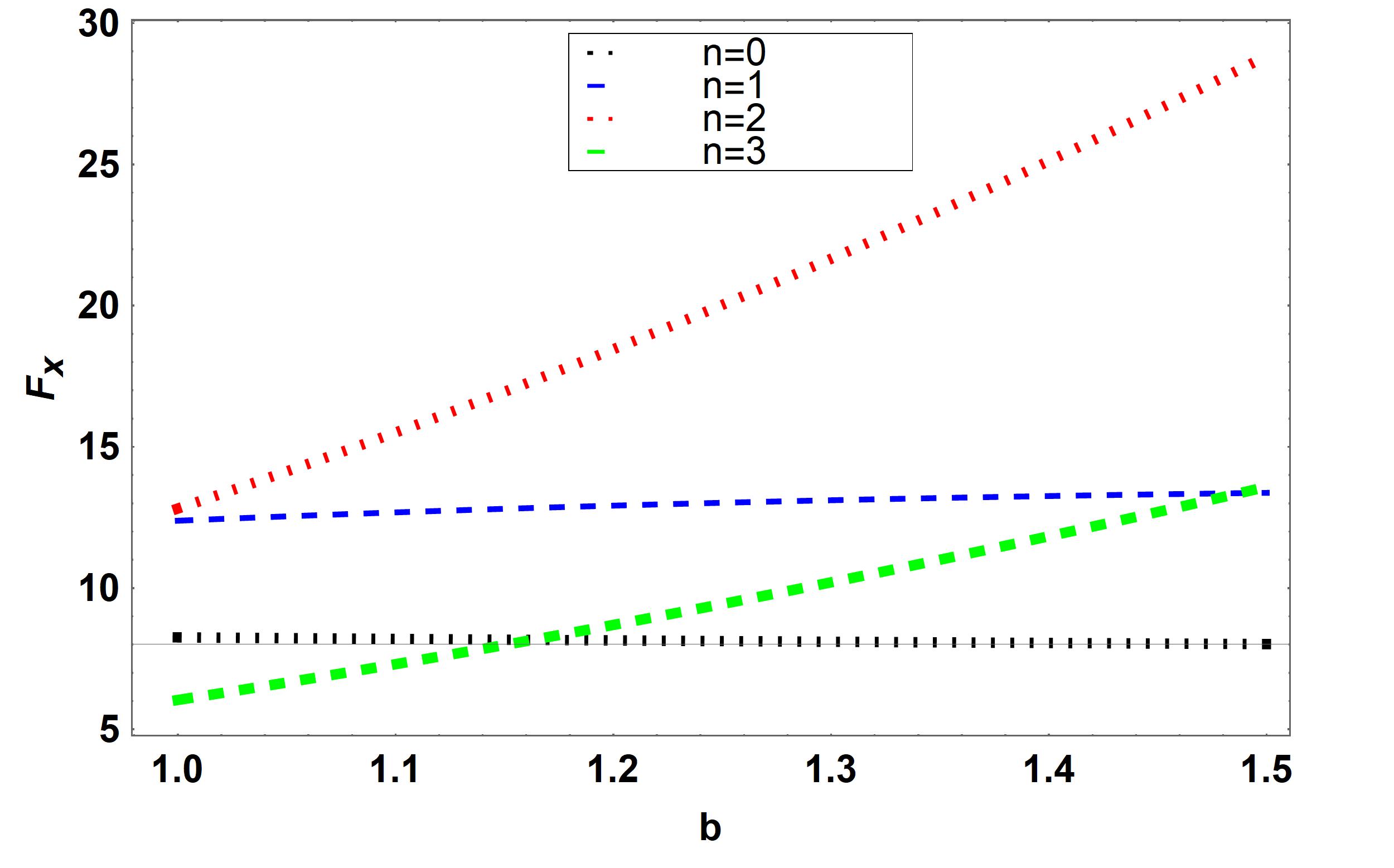} 
        \caption{Position space}
         \label{fig:image9}
    \end{subfigure}
    \hfill
    \begin{subfigure}[t]{0.45\textwidth}
        \centering
        \includegraphics[width=\textwidth]{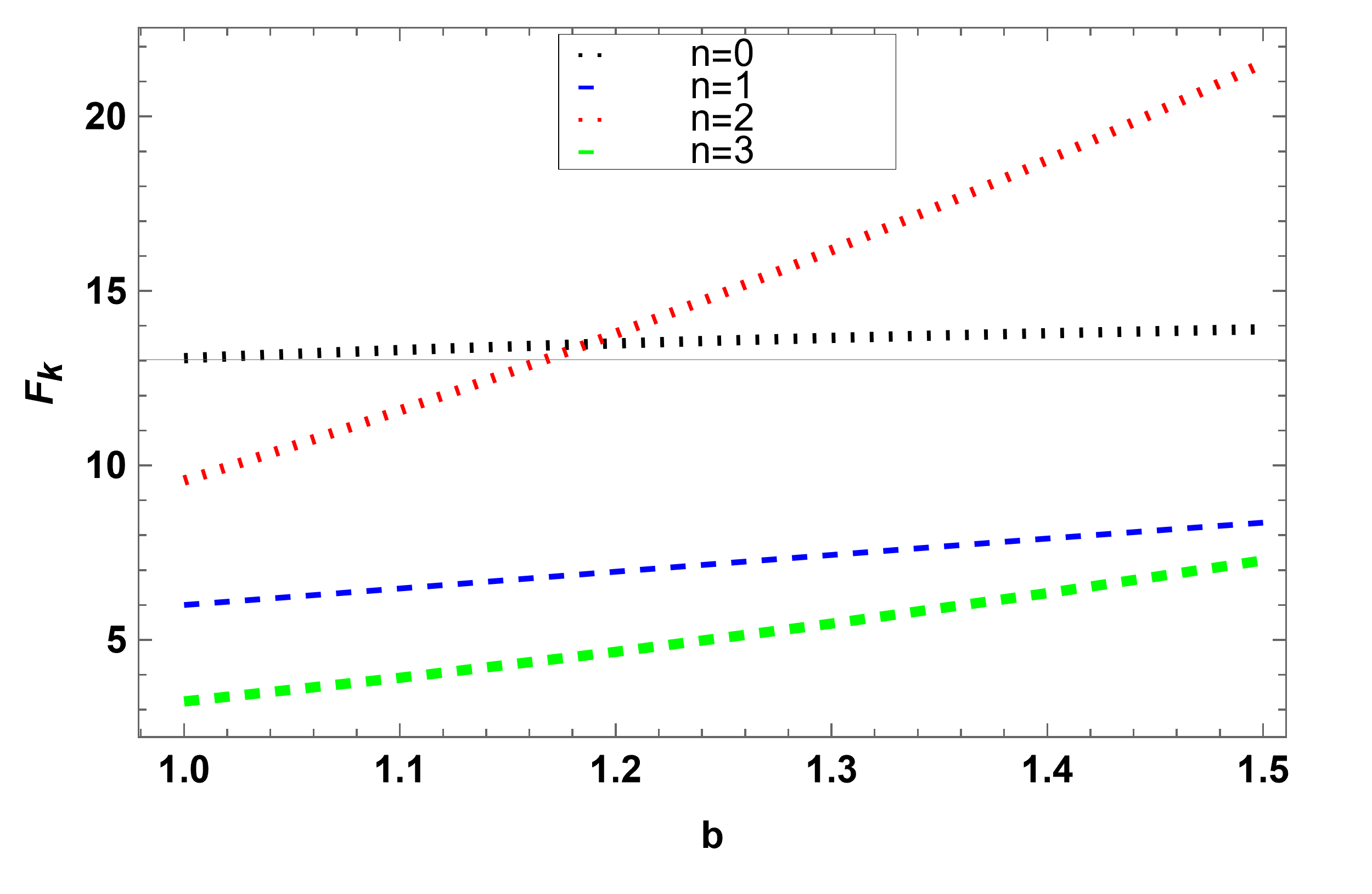} 
        \caption{Momentum space}
         \label{fig:image10}
    \end{subfigure}

    \vspace{1em} 
    \begin{subfigure}[t]{0.5\textwidth}
        \centering
        \includegraphics[width=\textwidth]{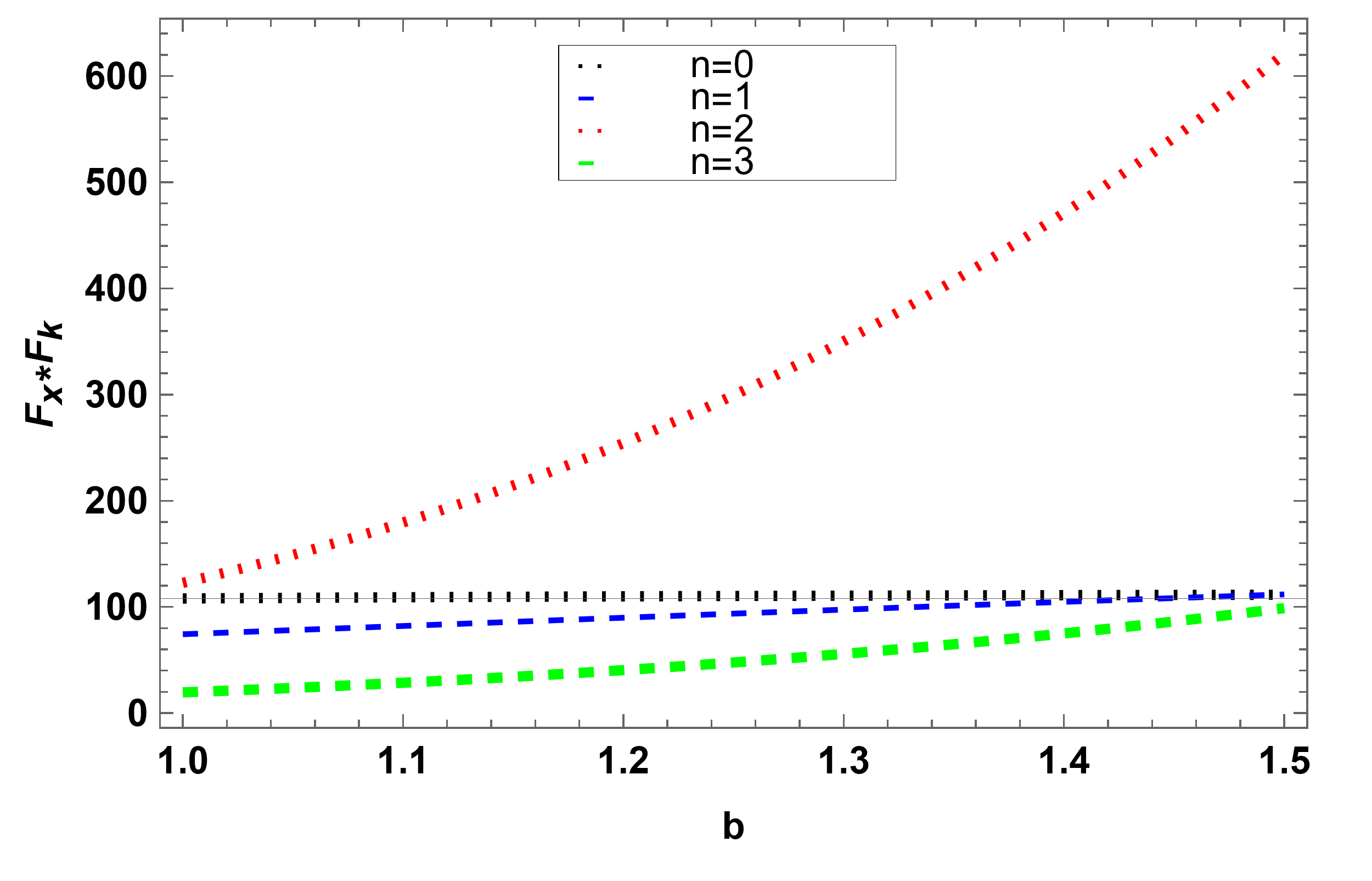} 
        \caption{Combined value of $F_x*F_k$}
         \label{fig:image11}
    \end{subfigure}

    \caption{Plots of Fisher's Information measure for the quartic potential.}
     \label{fig:image2main4}
\end{figure}

\begin{center}
\begin{table}[H]
    \begin{tabular}{|p{1cm}<{\centering}|p{3cm}<{\centering}|p{2cm}<{\centering}|p{2cm}<{\centering}|p{2cm}<{\centering}|p{2cm}<{\centering}|p{2cm}<{\centering}|}
        \hline
        \textbf{n} & \textbf{$\lambda$} & \textbf{$F_x$} & \textbf{$F_k$} & \textbf{$\sigma^{2}_{x}$} & \textbf{$\sigma^{2}_{k}$} & \textbf{$F_{x}\cdot F_{k}$}\\ \hline
        
        \multirow{3}{*}{0} & 1    & 0.751477 & 6.83279 & 1.70819  & 0.18786 & 5.1346\\ 
                             & 1.3    & 1.27 & 4.04307   & 1.10107     & 0.3175 & 5.1346 \\ 
                             & 1.5 & 1.69082 & 3.0368 & 0.7592   & 0.4227 & 5.13468 \\ \hline
                             
        \multirow{3}{*}{1} & 1    & 0.692013 & 89.116   & 22.279    & 0.1730 & 61.6694\\ 
                             & 1.3      & 0.899617 & 52.7314 & 13.1828   & 0.2249 & 47.4380\\ 
                             & 1.5   & 1.03802 & 39.6071   & 9.90177 & 0.2595 & 41.1129 \\ \hline
                             
        \multirow{3}{*}{2} & 1   & 3.7858 & 55.9765 & 13.9941         & 0.94645 & 211.9158 \\ 
                             & 1.3 & 6.398 & 33.1222 & 8.2805      & 1.5995 & 211.9158\\ 
                             & 1.5  & 8.51805 & 24.8784   & 6.2196    & 2.1295 & 211.9154 \\ \hline
                             
        \multirow{3}{*}{3} & 1  & 2.23831 & 41.1268 & 10.2817   & 0.5595 & 92.0545\\ 
                             & 1.3   & 3.78274 & 24.3354   & 6.0838 & 0.9456 & 92.0544 \\ 
                             & 1.5   & 5.03619 & 18.2786   & 4.56965 & 1.25904 & 92.0545\\ \hline

    \end{tabular}
    \caption{Numerical values of the Fisher's Information measure for symmetric well}
    \label{fishertable2}
    \end{table}
\end{center}

\begin{figure}[H]
    \centering
    \begin{subfigure}[t]{0.45\textwidth}
        \centering
        \includegraphics[width=\textwidth]{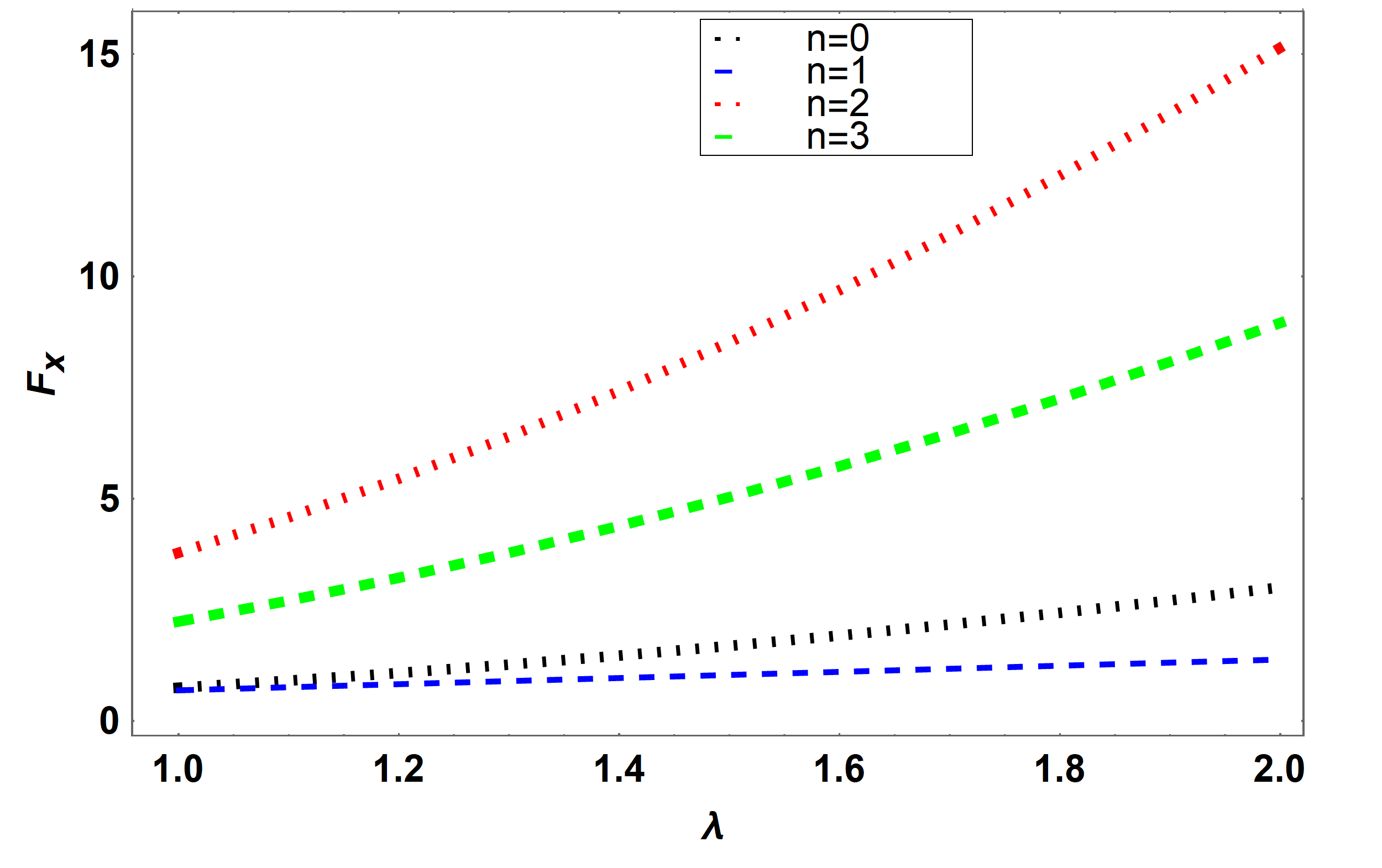} 
        \caption{Position space}
         \label{fig:image12}
    \end{subfigure}
    \hfill
    \begin{subfigure}[t]{0.45\textwidth}
        \centering
        \includegraphics[width=\textwidth]{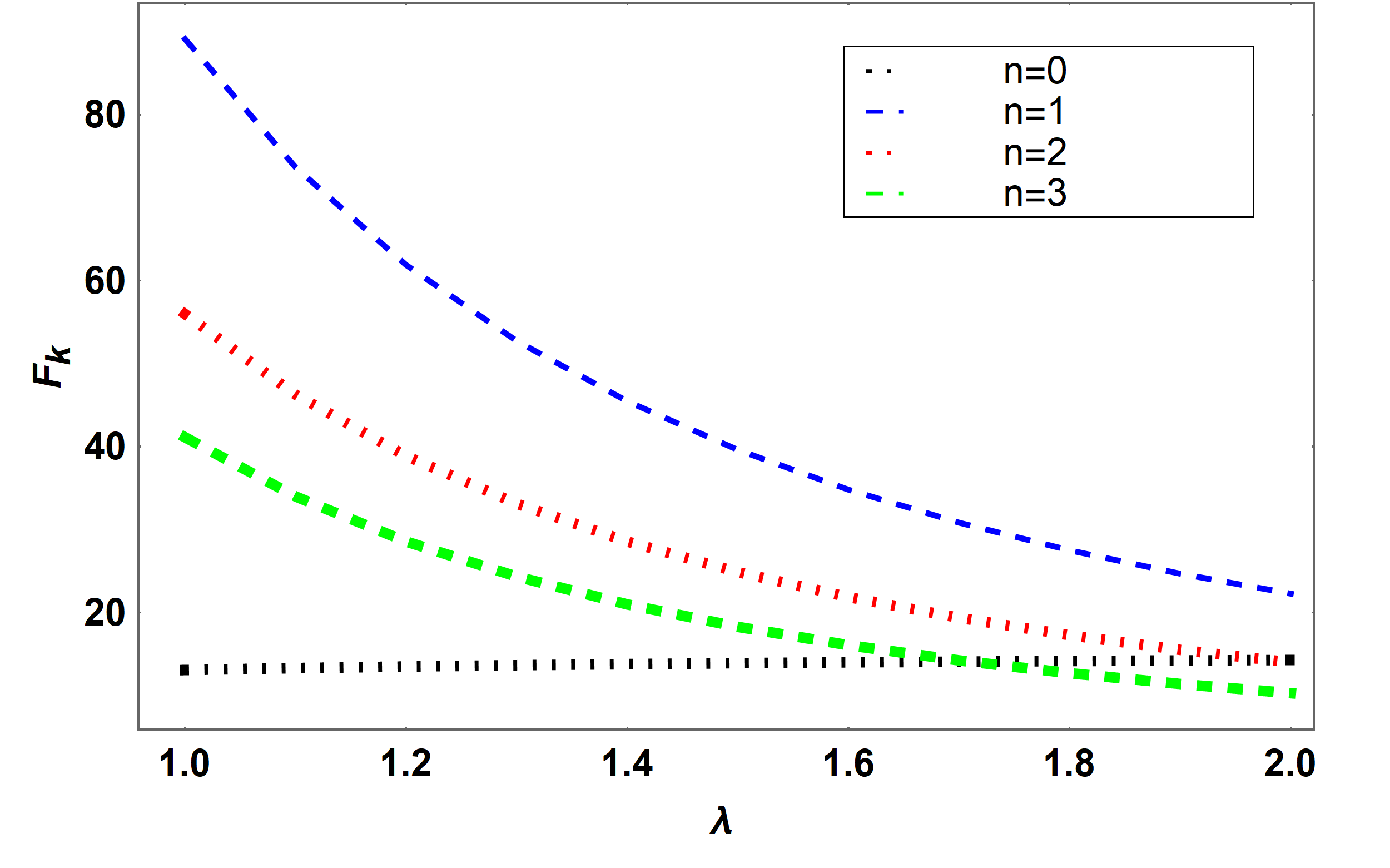} 
        \caption{Momentum space}
         \label{fig:image13}
    \end{subfigure}

    \vspace{1em} 
    \begin{subfigure}[t]{0.5\textwidth}
        \centering
        \includegraphics[width=\textwidth]{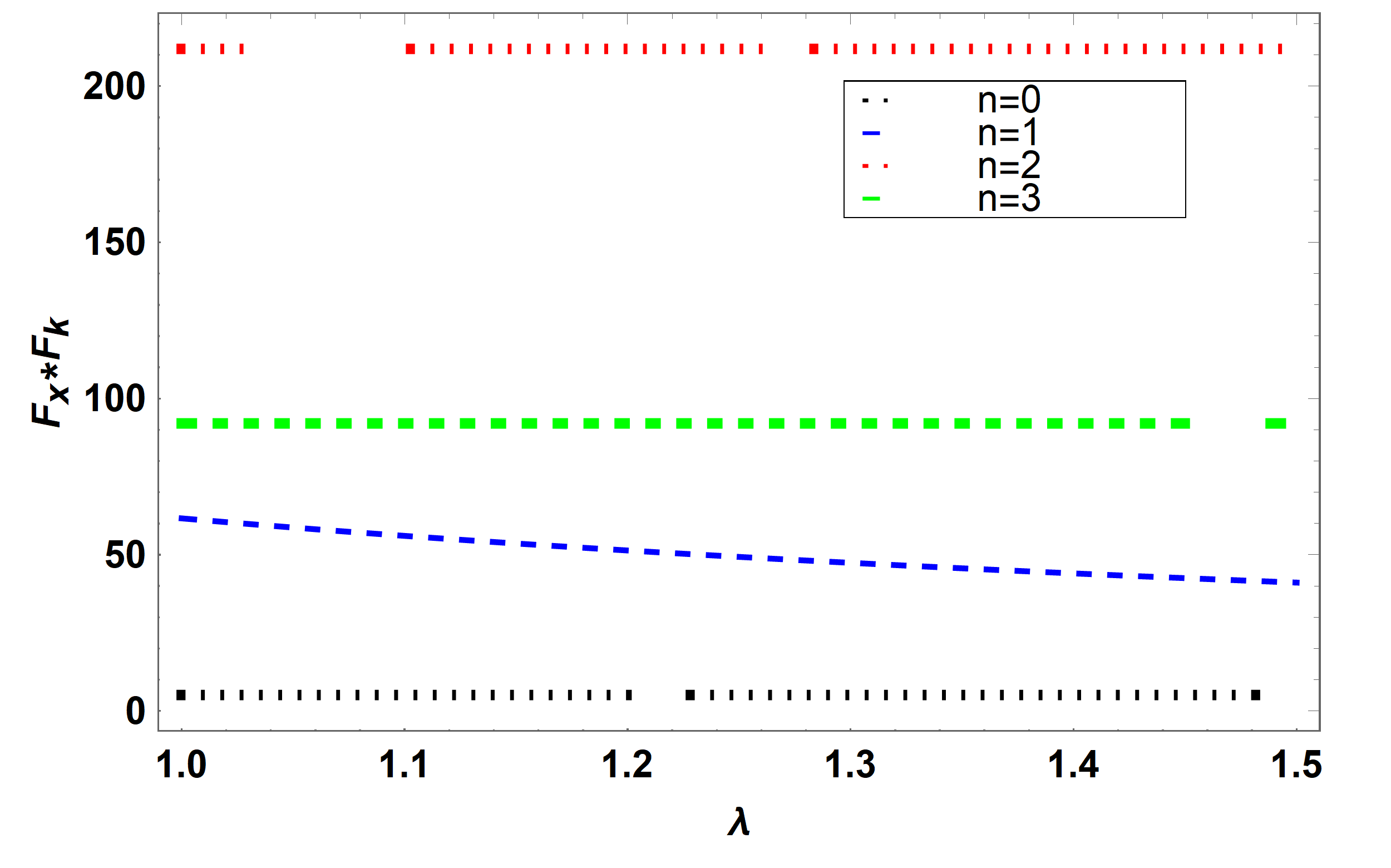} 
        \caption{Combined value of $F_x*F_k$}
         \label{fig:image14}
    \end{subfigure}

    \caption{Plots of Fisher's Information measure for the symmetric well.}
     \label{fig:image2main5}
\end{figure}

\section{Summary and Outlook} \label{sec:conclusion}
In this work, we have obtained the Shannon and Fisher information measures for systems subjected to quartic and symmetric well potentials. The solutions to the time-independent Schr\"{o}dinger equation were obtained for both systems using perturbation theory. After obtaining the solutions for both systems, we analyzed the Shannon entropy for the ground state, as well as the first, second, and third excited states, in relation to the width of the potential well ($b$ in case of quartic potential and $\lambda$ in case of symmetric well). Since we work within the regime of perturbation theory, a smaller perturbation parameter $(b)$ for the quartic potential and $(\lambda)$ for the symmetric well indicates a weaker interaction between the entities involved. Smaller values ensure that the series converges or remains asymptotically reliable. In systems with small perturbation parameters, nonlinear interactions (resulting from higher-order terms) contribute less to the overall dynamics. This simplifies the system, making it easier to analyze and predict behavior. Therefore, we take values of $(b = 1, 1.3, 1.5, \dots)$, and $(\lambda = 1, 1.3, 1.5, \dots)$ to preserve the regime in which we are working. Additionally, $(a = \frac{1}{2})$ in the quartic potential is chosen to ensure proper normalization.
We observe that, in the case of the quartic potential, the Shannon entropy decreases in position space while it increases in momentum space as the width $(b)$ increases. In case of symmetric well, the Shannon entropy decreases with the width $(\lambda)$ in position space and increases with the width $(\lambda)$ in momentum space. In both cases, this results in the sum of Shannon entropies in position and momentum space remaining constant with respect to the width of the potential., i.e.,
\begin{equation}
    \frac{d}{db}(S_{x}+S_{k}) = 0 = \frac{d}{d\lambda}(S_{x}+S_{k}).
\end{equation}
It can also be noted from the plots that the energy levels follow the same trend as the width of the potential well. From these plots, we deduce that the information contained in the system decreases as it becomes more localized in position space for both the quartic and symmetric well potentials. However, in momentum space, this localization leads to an increase in entropy, thereby increasing the uncertainty in momentum measurements. This behavior is due to two main factors: the symmetry of the potential well and the error corrections introduced by perturbation theory. Moreover, for each state, the value of $S_{x}+S_{k}$ remains constant which is also a consequence of Heisenberg's uncertainty principle \cite{lima2022quantum}.

Increases or decreases in Shannon entropy reflect how the system redistributes its uncertainty or disorder, influenced by physical processes such as diffusion, confinement, temperature changes, or quantum evolution. These can be visualized through our plots.

Upon analyzing the plots of Fisher information measures for the quartic potential, we observe that the information in both position and momentum space does not change significantly; it remains almost constant with some minor errors. This is not the case with the symmetric well, where Fisher information decreases as localization increases in position space but increases as localization increases in reciprocal space (momentum). We also observe from another plot an interesting feature of Heisenberg's uncertainty principle in the context of Fisher information measure, namely, $F_{x} = 4\sigma_{k}^{2}$ and $F_{k} = 4\sigma_{x}^{2}$. This results in $F_{x}\cdot F_{k}\geq 4\hbar^{2}$, which is also a consequence of the uncertainty principle. We can also conclude from these systems that, with greater localization of the potential well, less information content is transmitted; i.e., as position uncertainty decreases, the system's momentum uncertainty increases.

The plots can be analyzed using the Boltzmann $H$-theorem \cite{ben2017entropy,lesovik2016h}, which describes how entropy evolves in a system of particles as it moves towards equilibrium. This theorem introduces the concept of entropy (through the $H$-function) and explains the tendency of systems to evolve towards a state of maximum entropy. An increase in Fisher information in position space indicates that the probability density in position space becomes more localized. This suggests the system may be moving away from equilibrium in position space. Such localization could correspond to a lower-entropy state, as greater localization implies a more ordered system. In the context of the $H$-theorem, this behavior is analogous to a system influenced by kinetic or external forces driving it toward a non equilibrium state in position space. Conversely, a decrease in Fisher information in momentum space, reflecting a more spread-out distribution, suggests higher uncertainty in momentum. This typically corresponds to an increase in entropy in momentum space, aligning with the $H$-theorem's prediction that systems naturally evolve towards states of higher entropy. The reduction in Fisher information in momentum space could indicate the system redistributing momentum among particles to achieve equilibrium, consistent with the Boltzmann description.

In summary, the interplay between increasing Fisher information in position space and decreasing Fisher information in momentum space reflects a localized shift in the system's probability densities. Within the framework of the Boltzmann $H$-theorem, this can be understood as different facets of the system's approach to thermodynamic equilibrium, or as the impact of external perturbations driving it away from equilibrium in one domain while compensating in another.

Our future work investigates Shannon entropy and Fisher information for position-dependent mass distributions, exploring how localization affects information content. We aim to use Wigner and Husimi distributions \cite{radhakrishnan2022wigner,phasespace} to compare probability distributions for communication efficiency. Additionally, we'll study solitons in Bose-Einstein condensates within the Gross-Pitaevskii equation \cite{astrakharchik2011condensate,bradley2022scaling,cormack2012finite}, incorporating quantum coherence. This will highlight the universality of information entropy in characterizing soliton dynamics across systems.  \\ \\ 
\textit{Declaration:} All authors have equal contributions.

\appendix
\section{Wave function calculation for quartic potential}\label{sec:appendix1}
The quartic potential with parameters $a$ and $b$ is expressed as:
\begin{equation}
    V(x) = ax^2 + bx^4.
\end{equation}
We solve the Schrödinger equation to obtain the wave function. The Hamiltonian can be expressed as:
\begin{equation}
    H \psi = \left(\frac{p^2}{2} + ax^2 + bx^4\right)\psi.
\end{equation}

For simplicity, we take $a = \frac{1}{2}$. The Hamiltonian then becomes:
\begin{equation}
    H \psi = \left(\frac{p^2}{2} + \frac{1}{2}x^2 + bx^4\right)\psi.
\end{equation}

The Hamiltonian is expressed as a sum of the unperturbed and perturbed parts:
\begin{equation}
    H = H_0 + \lambda H',
\end{equation}
where $\lambda$ is the perturbation parameter. Let $H_0 = \frac{p^2}{2} + \frac{x^2}{2}$ and $H' = x^4$. We use operator formalism to find the wave function. Define the operators:
\begin{equation}
    a = \frac{1}{\sqrt{2}}(x + ip), \quad a^\dagger = \frac{1}{\sqrt{2}}(x - ip),
\end{equation}
which yield:
\begin{equation}
    x = \sqrt{2}(a + a^\dagger), \quad p = i\sqrt{2}(a - a^\dagger).
\end{equation}

The energy eigenvalues of the unperturbed Hamiltonian are:
\begin{equation}
    E_n^{(0)} = \left(n + \frac{1}{2}\right),
\end{equation}
where $n$ is the number operator, $n = a^\dagger a$. For the unperturbed system, the ground-state wave function is given by:
\begin{equation}
    \psi^{(0)}_n(x) = \frac{1}{\sqrt{2^n n!}}\left(\frac{1}{\pi}\right)^{1/4} H_n(x) e^{-x^2/2},
\end{equation}
where $H_n(x)$ is the Hermite polynomial. For the ground state ($n = 0$), this simplifies to:
\begin{equation}
    \psi^{(0)}_0(x) = \left(\frac{1}{\pi}\right)^{1/4} e^{-x^2/2}.
\end{equation}

The ground-state energy is:
\begin{equation}
    E_0^{(0)} = \frac{1}{2}.
\end{equation}

We now consider $H' = x^4$ as a perturbation. The first-order perturbed energy is given by:
\begin{equation}
    E_n^{(1)} = \langle \psi_n^{(0)} | H' | \psi_n^{(0)} \rangle.
\end{equation}

For the ground state ($n = 0$), substituting $H' = x^4$, we get:
\begin{equation}
    E_0^{(1)} = b \langle 0 | (a + a^\dagger)^4 | 0 \rangle.
\end{equation}

Expanding $(a + a^\dagger)^4$:
\begin{equation}
    (a + a^\dagger)^4 = a^4 + (a^\dagger)^4 + 4a^\dagger a + 6(a^\dagger a)^2 + \text{other terms}.
\end{equation}

Evaluating the expectation value:
\begin{equation}
    E_0^{(1)} = 3b.
\end{equation}

The total ground-state energy, including the perturbation, is:
\begin{equation}
    E_0 = E_0^{(0)} + E_0^{(1)} = \frac{1}{2} + 3b = \frac{6b + 1}{2}.
\end{equation}
The total wave function is:
\begin{equation}
    \psi_0(x) = \psi_0^{(0)}(x) + \psi_0^{(1)}(x),
\end{equation}
where the first-order correction $\psi_0^{(1)}(x)$ is:
\begin{equation}
    \psi_0^{(1)}(x) = \sum_{m \neq 0} \frac{\langle m^{(0)} | H' | 0^{(0)} \rangle}{E_0^{(0)} - E_m^{(0)}} | m^{(0)} \rangle.
\end{equation}

For $H' = x^4$, the correction simplifies to:
\begin{align}
    \psi_0^{(1)}(x) &= b \sum_{m \neq 0} \frac{\langle m | x^4 | 0 \rangle}{-m} | m \rangle \\
    &= b \Bigg[\frac{2\sqrt{2}\sqrt{3}|4\rangle}{-4} + \frac{\sqrt{2}|2\rangle}{-2} + \cdots \Bigg].
\end{align}

The total normalized wave function in position space is:
\begin{equation}
    \psi_0(x) = \left(\frac{1}{\pi}\right)^{1/4} \left[1 - b\left(\frac{1}{\sqrt{16 \cdot 4!}} \sqrt{\frac{3}{2}} (16x^4 - 48x^2 + 12) + \frac{3\sqrt{2}}{\sqrt{4 \cdot 2!}} (4x^2 - 2)\right)\right] e^{-x^2/2}.
\end{equation}
and in reciprocal space, the wave function is obtained by taking the Fourier transform,
\begin{equation}
    \phi(k)=0.187781 \bigg(\lambda (-4k^4+36k^2-15)+4\bigg)e^{\frac{-k^2}{2}}
\end{equation}
Upon normalizing the wave function in position space we get
\begin{equation}
    \psi_{0}(x) = \frac{1}{\sqrt{1+6.681\times 10^{-16}b+18.5b^{2}}}\bigg(\frac{1}{\pi}\bigg)^{1/4}\Bigg[1-b\bigg(\frac{1}{\sqrt{16.4!}}\sqrt{\frac{3}{2}}(16x^4-48x^2+12)+\frac{1}{\sqrt{4.2!}}3\sqrt{2}(4x^2-2)\bigg)\Bigg]e^{\frac{-x^2}{2}}
\end{equation}
Normalizing wave function in momentum space we get
\begin{equation}
    \phi_{0}(k) = \frac{0.0692}{\sqrt{1+6.681\times 10^{-16}b+18.5b^{2}}}\Bigg[12+b\bigg(-3(12+\sqrt{3})+12(6+\sqrt{3})k^{2}-4\sqrt{3}k^{4}\bigg)\Bigg]e^{\frac{-k^2}{2}}
\end{equation}
Wave function in the first excited state ($n=1$), 
\begin{equation}
   \psi_{1}(x) = A_{1}\Bigg[1-\frac{b}{32}\bigg\{10(2x^{2}-3)+\frac{1}{2}(4x^{4}-20x^{2}+15)\bigg\}\Bigg]xe^{-\frac{x^2}{2}},
\end{equation}
where $A_{1} = \frac{\sqrt{2}}{\pi^{1/4}\sqrt{1+2.505\times 10^{-16}b+0.6152 b^{2}}}$. In momentum space
\begin{equation}
\phi_{1}(k) = B_{1}\Bigg[4k^{4}b-60k^{2}b+75 b-64\Bigg]k e^{-\frac{k^{2}}{2}},
\end{equation}
where $B_{1} = \frac{-0.0165 i}{\sqrt{1+2.505\times 10^{-16}b+0.6152 b^{2}}}\sqrt{\frac{1.625 +4.071\times 10^{-16}b+b^{2}}{1.606+8.049\times 10^{-16}b + 0.9883b^{2}}}$.
The wave function for $n=2$ we get
\begin{equation}
    \psi_{2} = \frac{b}{16}\bigg[12\sqrt{2}\ket{0}-28\sqrt{12}\ket{4}-6\sqrt{10}\ket{6}\bigg]
\end{equation}
The normalized wave function for $n=2$ is given by
\begin{equation}
    \psi_{2}(x) = \frac{128}{12183}\frac{b}{16}\bigg(\frac{1}{\pi}\bigg)^{\frac{1}{4}}\bigg[12\sqrt{2}-14\sqrt{2}(4x^4-12x^2+3)-\frac{1}{\sqrt{2}}(8x^6-60x^4-90x^2-15)\bigg]e^{-\frac{x^2}{2}}
\end{equation}

The normalized wave function in momentum space is 

\begin{equation}
   \phi_{2}(k)=\frac{b e^{\frac{-k^2}{2}}(105 + 246 k^2 - 172 k^4 + 8 k^6)}{2 \sqrt{12183} \pi^{1/4}} 
\end{equation}
\newpage

\section{Wave function calculation for symmetric well}\label{sec:appendix}
The symmetric well with parameter $\lambda$ can be written as
\begin{equation}
    V(x) = V_{0}\bigg(\frac{1-\lambda x \cot (\lambda x)}{(\lambda x)^{2}}\bigg).
\end{equation}
The time-independent Schr\"{o}dinger equation is given by
\begin{equation}
    H\psi = \frac{-1}{2m}\frac{d^2\psi(x)}{dx^2} + V_{0}\bigg(\frac{1-\lambda x \cot(\lambda x)}{(\lambda x)^{2}}\bigg) \psi (x).
\end{equation}
We use stationary perturbation theory as in ref. \cite{carvalho2019solution}) to obtain this system's wave function and energy levels. This potential (up to $\mathcal{O}(x^{6})$ term) is expressed as
\begin{equation}
    V(x) \approx V_{0}\bigg(\frac{1}{3}+\frac{1}{45}(\lambda x)^{2}+\frac{2}{945}(\lambda x)^{4}+\frac{1}{4725}(\lambda x)^{6}\bigg).
\end{equation}
To solve this equation we do a transformation $\chi = \lambda x$ and we consider $V_{0} = 1$. 
\begin{equation}
    -\frac{d^2\phi(\chi)}{d\chi^{2}}+\frac{1}{2}\Omega^{2}\chi^{2}\phi(\chi) + \bigg[\frac{1}{3}+\frac{2\chi^{4}}{945}+\frac{\chi^{6}}{4725}\bigg]\phi(\chi) = \epsilon\phi(\chi),
\end{equation}
where $\Omega = \sqrt{\frac{2}{45}}$ and $\epsilon = \frac{mE}{\hbar^{2}\lambda^{2}}$. We use creation and annihilation operators to solve this problem.
\begin{equation}
    a^{\dagger} = \sqrt{\frac{\Omega}{2}}\bigg(\chi - \frac{ip}{\Omega}\bigg), \hspace{2em} a = \sqrt{\frac{\Omega}{2}}\bigg(\chi + \frac{ip}{\Omega}\bigg) \implies \chi = \sqrt{\frac{1}{2\Omega}}(a+a^{\dagger}),
\end{equation}
where$$
    a\ket{n} = \sqrt{n}\ket{n-1} \hspace{2em} a^{\dagger}\ket{n} = \sqrt{n+1}\ket{n+1}.$$
According to stationary perturbation theory, the first-order correction to the energy simply equals the mean
value of perturbation term $H^{\prime}$ in the unperturbed state $\ket{n}$. 
\begin{equation}
    \epsilon^{(1)}_{n} =  \bra{n}H^{\prime}\ket{n}.
\end{equation}
Energy in the first excited state ($n=1$)
\begin{equation} 
    \epsilon^{(1)}_{n} = \frac{1}{3} + \frac{2\chi^{4}}{945}\bigg(\frac{1}{2\Omega}\bigg)^{2} (6n^2+6n+3) + \frac{1}{4725} \bigg(\frac{1}{2\Omega}\bigg)^{2} (20n^{3}+30n^{2}+40n+15),
\end{equation}
where $\Omega^{2} = \frac{2}{45}$.
\begin{equation}
    \epsilon^{(1)}_{n} = \frac{1}{3} + \frac{1}{84}(6n^{2}+6n+3) + \sqrt{\frac{45}{2}}\bigg(\frac{1}{1680}\bigg)(20n^3+30 n^{2} + 40 n + 15).
\end{equation}
The second-order perturbation energy correction is given 
\begin{equation}
    \epsilon^{(2)}_{n} = \sum_{n\neq m} \frac{|\bra{n}H^{\prime}_{\chi^{4}}|\ket{m}|^{2}}{\epsilon^{(0)}_{m}-\epsilon^{(0)}_{n}}.
\end{equation}
At ground state energy correction at second order is equal to $-0.02823$. The total energy is given by
\begin{equation}
    \epsilon_{n} = \epsilon^{(0)}_{n}+\epsilon^{(1)}_{n}+\epsilon^{(2)}_{n}.
\end{equation}
The wave function is obtained as follows 
\begin{equation}
   \psi = \psi^{(1)}_{n}(r) = \psi^{(0)}_{n}(r) + \sum_{n\neq k} \frac{\bra{\psi^{(0)}_{k}}H^{\prime}\ket{\psi^{(0)}_{n}}}{E^{(0)}_{n}-E^{(0)}_{k}} \ket{\psi^{(0)}_{k}},
\end{equation}
\begin{equation}
    \psi^{(0)}_{n}(r) = \bigg(\frac{m\Omega}{\pi \hbar}\bigg)^{1/4} \frac{1}{\sqrt{2^{n}n!}} H_{n}(r) e^{\frac{-r^2}{2}},
\end{equation}
where $r = \sqrt{\frac{m\Omega}{\hbar}}r = \bigg(\frac{2}{45}\bigg)^{1/4} \chi$ and $H_{n}(r)$ is Hermite polynomial \cite{arfken2011mathematical}.
\begin{equation}
    \psi^{(1)}_{0}(r) = \psi^{(0)}_{0}+\frac{\bra{\psi^{(0)}_{2}}H^{\prime}_{\chi^4}\ket{\psi^{(0)}_{0}}}{\epsilon^{(0)}_{0}-\epsilon^{(2)}_{0}}\psi_{2}^{(0)}+\frac{\bra{\psi^{(0)}_{4}}H^{\prime}_{\chi^4}\ket{\psi^{(0)}_{0}}}{\epsilon^{(0)}_{0}-\epsilon^{(0)}_{4}}\psi_{4}^{(0)}.
\end{equation}
Thus the ground state wave function of the system is given by
\begin{equation}
    \psi(r) = \bigg(\frac{1}{\pi}\sqrt{\frac{2}{45}}\bigg)^{1/4} e^{\frac{-r^2}{2}}\bigg[1-0.0847(4r^{2}-2)-0.0035(16r^{4}-48r^{2}+12)\bigg],
\end{equation}
\begin{equation}
    \psi(x) = 0.9703\sqrt{\lambda} \bigg(\frac{1}{\pi}\sqrt{\frac{2}{45}}\bigg)^{1/4} e^{\frac{-\lambda^{2}x^{2}}{3\sqrt{10}}}\Bigg[1-0.1694\bigg(\frac{2\sqrt{2}}{\sqrt{45}}\lambda^{2}x^{2}-1\bigg)-0.014\bigg(\frac{8}{45}\lambda^{4}x^{4}-12\sqrt{\frac{2}{45}}\lambda^{2}x^{2}+3\bigg)\Bigg].
\end{equation}
We extend the same process for $n=2$,
\begin{equation}
    \psi^{(1)}_{2}(x) = \psi^{(0)}_{2}(x) + \sum_{k\neq 2} \frac{\bra{\psi^{(0)}_{k}}H^{\prime}\ket{2^{(0)}}}{\epsilon^{(0)}_{2}-\epsilon^{(0)}_{k}},
\end{equation}
and we get the wave function as
\begin{multline}
    \psi(x) (\text{for $n=2$}) = \ket{2} -\frac{2}{945} \frac{1}{4\Omega^{3}}\Bigg\{-\frac{3}{2}\sqrt{10}\ket{6}-8\sqrt{35}\ket{8}+3\sqrt{2}\ket{0}-8\sqrt{15}\ket{4}-2\sqrt{3}\ket{4}\Bigg\} + \frac{1}{4725}\frac{1}{8\Omega^{4}}\\\Bigg\{-4\sqrt{35}\ket{8}-\frac{81}{4}\sqrt{10}\ket{6}+\frac{45}{\sqrt{2}}\ket{0}-195\sqrt{3}\ket{4}\Bigg\},
\end{multline}
where $r = \sqrt{\frac{m\Omega}{\hbar}}r = \bigg(\frac{2}{45}\bigg)^{1/4} \chi$ and $H_{n}(r)$ is Hermite polynomial. 
\bibliographystyle{plain}
\bibliography{ref}
\end{document}